\documentclass[a4paper,fleqn,usenatbib]{mnras}

\usepackage{newtxtext,newtxmath}

\usepackage[T1]{fontenc}
\usepackage{ae,aecompl}

\usepackage{graphicx}	
\usepackage{amsmath}	
\usepackage{subfigure}
\usepackage{longtable}

\title[Dense gas in galaxies]{Dense gas in local galaxies revealed by multiple tracers}

\author[Fei Li et al.]{Fei Li$^{1}$, 
Junzhi Wang$^{1,2}$\thanks{E-mail: jzwang@shao.ac.cn},  Feng Gao$^{3}$, Shu Liu$^{4}$, Zhi-Yu Zhang$^{5}$, Shanghuo Li$^{1,6}$
\newauthor Yan Gong$^{7}$, Juan Li$^{1,2}$ and Yong Shi$^{5}$
\\
$^{1}$Shanghai Astronomical Observatory, Chinese Academy of Sciences,80 Nandan Road, Shanghai, 200030, China\\
$^{2}$Key Laboratory of Radio Astronomy, Chinese Academy of Sciences,  10 Yuanhua Road, Nanjing, JiangSu 210033, China\\
$^{3}$Max-Planck-Institut für Extraterrestrische Physik, Gie{\ss}enbachstrasse 1, D-85741 Garching bei München, Germany\\
$^{4}$CAS Key Laboratory of FAST, National Astronomical Observatories, Chinese Academy of Sciences, Beijing 100012, China\\
$^{5}$School  of Astronomy and Space Science, Nanjing University, Nanjing,  210093, China\\
$^{6}$Korea Astronomy and Space Science Institute, 776 Daedeokdae-ro, Yuseong-gu, Daejeon 34055, Republic of Korea \\
$^{7}$Max-Planck-Institut für Radioastronomie, Auf dem Hügel 69, 53121, Bonn, Germany \\
}

\date{Accepted XXX. Received YYY; in original form ZZZ}

\pubyear{2020}

\begin{document}
\label{firstpage}
\pagerange{\pageref{firstpage}--\pageref{lastpage}}
\maketitle

\begin{abstract}
We present  3 mm and 2 mm band simultaneously  spectroscopic observations of HCN 1-0, HCO$^+$ 1-0, HNC 1-0, and CS 3-2 with the IRAM 30 meter telescope,  toward a sample of 70 sources as nearby galaxies with infrared luminosities ranging from   several 10$^{5}L_{\odot}$ to more than 10$^{12}L_{\odot}$. After combining HCN 1-0, HCO$^+$ 1-0 and  HNC 1-0 data from literature  with our detections,  relations  between  luminosities of dense gas tracers (HCN 1-0, HCO$^+$ 1-0 and  HNC 1-0) and infrared luminosities are derived, with tight linear correlations  for all tracers.  Luminosities of CS 3-2 with only our observations also show tight linear correlation with infrared luminosities. No systematic difference is found for tracing dense molecular gas  among these tracers. Star formation efficiencies for dense gas with different tracers also do not show any trend  along  different   infrared luminosities. Our study also shows that HCN/HCO$^+$ line ratio might not be a good indicator to diagnose obscured AGN in galaxies.

\end{abstract}


\section{Introduction}

Dense gas tracers,  lines from molecules with high dipole moment (e.g., HCN,  HCO$^+$, HNC, and CS), which have high critical densities ($n\rm_{crit}$) greater than $10^4$ cm$^{-3}$, directly represent molecular content involved in forming stars \citep{2012ApJ...745..190L}. 
 Observing  star-forming molecular gas is one of the direct ways to measure  star formation  activities in galaxies and in the Milky Way. The star formation law, the relation between star formation rate (SFR) and local gas density, was first proposed by \cite{1959ApJ...129..243S} and observationally studied with  total gas, including atomic gas traced by HI 21 cm emission and molecular gas traced by CO 1-0 emission,  and SFR by  \cite{1998ApJ...498..541K}. With observations of HCN 1-0 toward 65 galaxies,  \cite{2004ApJ...606..271G}  found a tight linear correlation between the luminosities of HCN 1-0 and infrared emission. This relation has been extended to Galactic dense cores \citep{2005ApJ...635L.173W}  and possibly high-z galaxies and QSOs \citep{2007ApJ...660L..93G}.

However, since there are several choices of molecules (HCN, HNC, HCO$^+$ and CS) with high dipole moments, different masses of dense gas  will be obtained with different tracers, with conversion factor of line luminosity to gas mass in galaxies. It was argued  if HCN was a true  tracer of dense molecular gas in  luminous infrared galaxies (LIRGs) and ultra-luminous infrared galaxies (ULIRGs)  \citep{2006ApJ...640L.135G}. Further observational results in literature indicated that  HCO$^+$/HCN 1-0 line ratio varied from 0.36 to 1.83, while HNC/HCN 1-0 ratios were from 0.20 to 1.09  in a sample of  23 galaxies \citep{2011A&A...528A..30C}, most of which are (U)LIRGs.  Thus, there will be large uncertainties of estimating dense gas masses in individual galaxies  with only line luminosity of  one dense gas  tracer, even though  the correlations between line luminosities of deferent dense gas tracers and IR luminosities  have been well established
  \citep{2004ApJ...606..271G,2011MNRAS.416L..21W,2014ApJ...784L..31Z}.  Observations  of multiple dense gas tracers  toward  a relatively large sample up to about one hundred   of galaxies are essential for determining the uncertainty of estimating dense gas masses there. With these lines detected in galaxies, one can also determine the line ratios between each of them among different types of galaxies: with active galactic nucleus (AGN)
  or without AGN, starburst or normal star forming galaxies. 
 
The other way  for better understanding dense gas properties is to detect optically thin isotopic lines of dense gas tracer with deep observations \citep{2014ApJ...796...57W,2016MNRAS.455.3986W,2020MNRAS.494.1095L}.  The  isotopic lines of dense gas tracers together with dense gas tracers themselves, can be used to derive optical depths of dense gas tracers, which allows us to constrain dense gas masses more accurately than without isotopic lines.  The optical depths of HCN 1-0 derived with HCN/H$^{13}$CN 1-0 line ratios   vary from  less than 1 to more than 10 in a limited sample of 5 galaxies \citep{2020MNRAS.494.1095L}.  Such variation can cause large uncertainty of estimating dense gas masses with line luminosity of HCN 1-0 in galaxies.     
However, due to the weak intensities of these isotopic lines, only extremely local star forming galaxies or nearby extremely active starburst galaxies with strong line emission can be done for such study. Thus,  it is almost impossible to detect isotopic lines of dense gas tracers toward a large sample of galaxies with a reasonable amount of telescope time.  
Observations of multiple dense gas tracers toward different types of galaxies with infrared luminosities spanning several orders of magnitude will be the best way to testify the consistence of different dense gas tracers in galaxies.

In this paper, we will describe the sample in \S2, observation and data reduction  in \S3,  present the main results   in \S4 and discussions in \S5, make the brief summary  in \S6.

\section{The sample of galaxies}

We obtained new observations for a sample of  local galaxies including a wide range of infrared luminosities, from about several 10$^{5}L_{\odot}$ to more than 10$^{12}L_{\odot}$ spanning about 7 orders of magnitude. Firstly, 60 sources with bright infrared emission from the IRAS Revised Bright Galaxy Sample \citep{2003AJ....126.1607S} with DEC higher than -20$^{\circ}$ and $F_{100{\mu}m}>30Jy$ are selected.  Sources within that criteria which had been observed by \cite{2011A&A...528A..30C} are excluded. Since there are large fraction of  sources with $L_{IR}$  between 10$^{10}L_{\odot}$ and 10$^{11}L_{\odot}$,  we also excluded some sources within this range, especially  if their  LST ranges  are  good for Galactic plane observations. Other ten sources reach this  criteria observed with our previous projects are also included.  They are  six infrared bright galaxies: M~82 (center), NGC~3079, IC~694, Mrk~231, NGC~6240, and NGC~6946  from  \cite{2020MNRAS.494.1095L}, as well as  four different positions along the major axis of M~82  (Li et al. in preparation).  

\section{Observations and data reduction}
\subsection{Millimeter spectroscopic observations with IRAM 30m and data reduction}
The HCN 1-0, HCO$^+$ 1-0, HNC 1-0 and CS 3-2  lines were observed simultaneously with the IRAM 30 meter millimeter telescope (IRAM 30m) at
Pico Veleta, Spain on February, May, July and August 2019.   The critical densities  of  HCN 1-0, HCO$^+$ 1-0, HNC 1-0 and CS 3-2  are 6.8$\times10^4$cm$^{-3}$, 4.7$\times10^5$cm$^{-3}$, 1.4$\times10^5$cm$^{-3}$, and 4.4$\times10^5$cm$^{-3}$, when kinematic temperature of  H$_2$ is 10 K, while the effective excitation densities for these lines are about 1/50 of the critical densities  \citep{2015PASP..127..299S}.

We used the 3 mm (E0) and 2mm (E1) band of The Eight Mixer Receiver (EMIR) simultaneously and the  Fourier Transform Spectrometers (FTS) backend to cover 8 GHz bandwidth and 195 kHz spectral resolution for each band with dual polarization. Standard wobbler switching mode with beam throws of $\pm$60$^{''}$ and a
switching frequency of 0.5 Hz were used. The beam sizes of IRAM 30m  range from FWHM $\sim$ 28$^{''}$ at 88 GHz to $\sim$ 17$^{''}$ at 146 GHz. The typical system temperatures are around 100 K in the 3-mm band and around
150 K in the 2-mm band. The detailed informations of the sample are  shown in Table \ref{tab:observation parameters}, which includes source name, coordinates, IRAS 100${\mu}$m flux, $cz$ and luminosity distance \citep{2003AJ....126.1607S}, as well as observing dates, telescope time (ON+OFF) and typical system temperature at 3mm band.  Observations done in 2017 and 2019 Feb are with project 058-17 \citep{2020MNRAS.494.1095L} and 186-18  (Li et al. in preparation), respectively, while the rest observations are all  with project 066-19. 
The weather varies from good ($pwv \sim $3.8 mm) to mediocre ($pwv \sim$9.4 mm) conditions during the observations. Pointing was checked every 2 h with nearby strong millimetre emitting quasi-stellar objects. We also checked and corrected the focus at the beginning of each run and during sunsets and sunrises. The antenna temperature ($T_{\rm A}^{\ast}$) was converted to the main beam brightness temperature ($T_{\rm mb}$), using $T_{\rm mb}$=$T_{\rm A}^{\ast}\cdot F_{\rm eff}/B_{\rm eff}$,
where the forward efficiency $F_{\rm eff}$ is 0.95 and beam efficiency $B_{\rm eff}$ is 0.81 for 3 mm band, while  $F_{\rm eff}$ is 0.93 and   $B_{\rm eff}$ is 0.73 for 2 mm band.  Each scan consists of 2 minutes with an on-source integration of 1 minute.  
The total telescope time (ON+OFF) was about one hour towards each source.

Data reduction was conducted with the class package, which is a part of the GILDAS\footnote{http://www.iram.fr/IRAMFR/GILDAS} software. Firstly, each individual spectrum was checked. Then, we discarded the spectra with issues related to  their baseline flatness, standing wave, system temperature, etc. Most of the spectra have good qualities. All reliable spectra of a given source tuned at the same frequency were averaged into one spectrum. A first-order polynomial baseline was fitted and subtracted from the averaged spectrum.  Due to the broad line widths of these galaxies, the final spectra were smoothed to a velocity spacing of $\sim$20--40 km s$^{-1}$.   The rest frequencies of each line from the NIST database\footnote{https://pml.nist.gov/cgi-bin/micro/table5/start.pl} were used for the line identification.

\subsection{Infrared data and line luminosities}

To measure the total infrared luminosities  $L_{\rm IR}$ (3 $\mu$m - 1000$\mu$m), we obtained infrared archival imaging data from \emph{Spitzer} MIPS and \emph{Herschel} PACS instruments from the NASA/IPAC Infrared Science Archive (IRSA). The data have been processed to level 2 for MIPS 24 $\mu$m and level 2.5 or 3 for PACS 70 $\mu$m, 100 $\mu$m, and 160 $\mu$m bands. To match the molecular emission, the infrared luminosities  have been corrected from the whole galaxy to the region within the
IRAM 30-m telescope beam. Following the same approach as adopted by \cite{2018ApJ...860..165T}, \emph{Spitzer}  and \emph{Herschel} data have been convolved to 28$''$ and 17$''$ resolution, corresponding to the beam-size of the 3 mm and 2 mm band, respectively. For the galaxies with only \emph{Spitzer} MIPS 24 $\mu$m image, we used MIPS 24 $\mu$m image to obtain the ratio of such a region to the whole galaxy. Then, with the 24 $\mu$m flux ratio, we estimated the infrared luminosity within the beam by scaling the total infrared luminosity from \cite{2003AJ....126.1607S}, which is similar to the method in \cite{2011MNRAS.416L..21W}. We computed the  line luminosities for all dense gas tracers  using equation (2) in \cite{2004ApJ...606..271G} for all galaxies:

 \begin{equation}
\label{eq:pop}
L'_{\rm gas} \approx \pi/(4ln2)\theta^2I_{\rm HCN}d_L^2(1+z)^{-3}  {[\rm K~km~s^{-1}~pc^2]}  
\end{equation}

where I$_{\rm HCN}$ is the observed integrated line intensity, d$_L^2$ is the luminosity distance.

The infrared luminosity is calculated from

\begin{equation}
\label{eq:pop}
 L_{\rm TIR}=\Sigma c_{i}vL_{v}(i)L_{\odot}  
\end{equation}
where $c_{i}$ is the calibration coefficients for various combinations of bands, $vL_{v}(i)$ is the resolved luminosity in a given band $i$ in units of $L_{\odot}$  \citep{2018ApJ...860..165T}. The values of  $c_{i}$ are from \cite{2018ApJ...860..165T}.   The total error estimated for  $L_{\rm TIR}$ includes errors of photometry of $\sim 5\%$ \citep{2014ExA....37..129B}, the flux calibration error of $\sim 5\%$,  and the error of tracing TIR with a combined IR band of $\sim 20\%$ \citep{2013MNRAS.431.1956G} .

\section{Results}

\subsection{Detections of dense gas tracers in local galaxies with IRAM 30 meter}

The  targets with new observations, which include 70 sources, are  listed in Table \ref{tab:observation parameters}.  Dense gas tracers in these sources were detected  with detection rate of 89\%, 90\%, 73\%, and 51\%  for HCN 1-0, HCO$^+$ 1-0, HNC 1-0 and CS 3-2, respectively. The velocity-integrated intensities of these four lines are shown in Table \ref{tab:intensity}. For HCN 1-0 and HCO$^+$ 1-0, most of them are solid  detections  above 5 $\sigma$, while, some marginal detections at $\sim$3$\sigma$ to 4$\sigma$ level are shown for HNC 1-0 and CS 3-2.  3 $\sigma$ upper limits for the velocity-integrated intensity of the undetected lines are also derived.   The 3 $\sigma$ upper limits are calculated from $\sigma_{\rm line}$ = RMS$_{\rm channel}\sqrt{{\vartriangle}\mu\times {\delta}V}$, where RMS$_{\rm channel}$ is the baseline rms of the smoothed spectrum at  $\sim$  30 km/s-40 km/s and  ${\vartriangle}\mu$ is the line width and ${\delta}V$ is the channel width. For the sources of non-detection in all four species, including NGC~2403, NGC~4605, NGC~4654 and NGC~6822, we use the CO line width to estimate the upper limits of velocity-integrated  intensity, due to the central velocity of HCN 1-0 is consistent with that of the detected lines of CO 1-0 \citep{1995A&A...295..599I} and CO 3-2 \citep{2010ApJ...724.1336M}. For the other sources, we use line width of the detected lines to estimate the upper limits of velocity-integrated  intensity of the undetected lines for the same source.  
 Note that  CS 3-2  is not covered by the observed frequency range in  Mrk~231 and NGC~6240. The spectra of these dense gas tracers are shown in Figure \ref{fig:f1}, with the grey shaded region in each subfigure to indicate velocity range of identified  detections.

For comparison,  Arp 220, which was included in \cite{2015ApJ...814...39P}, is also observed. However, fluxes measured by our observations are only about 70\% of that  in \cite{2015ApJ...814...39P}.  Our fluxes are consistent with results reported in \cite{2009ApJ...692.1432G} and \cite{2016MNRAS.455.3986W}.  We speculate the mis-match of line fluxes of   Arp 220  in \cite{2015ApJ...814...39P} may be due to   poor velocity resolution, which can cause un-resolved absorption dip at central velocity \citep{2009ApJ...692.1432G,2016MNRAS.455.3986W}, or poor baselines, which can cause over-estimation of line widths.  Another evidence that such different fluxes  can not be caused by the pointing error of our observation is that the peak emissions are similar in all observations \citep{2009ApJ...692.1432G,2015ApJ...814...39P,2016MNRAS.455.3986W}, which is about 17mK  ($T_{\rm mb}$) for HCN 1-0 (see Figure \ref{fig:f1}).  Since line widths in  most of  other sources in  \cite{2015ApJ...814...39P} are not as broad as that in Arp 220 and absorption dip is not as important as that in Arp 220, such mis-match of line fluxes in \cite{2015ApJ...814...39P} should not be a big issue. 

Mrk 231, NGC 3079, NGC 4194, NGC 4418 and NGC 6240 were also included both in \cite{2011A&A...528A..30C} and our observations, with similar velocity integrated fluxes.  Results from our observations are used in discussion. 
By combining our new IRAM data with literature data, including 57 local (U)LIRGs \citep{2015ApJ...814...39P}  and 12 local galaxies \citep{2011A&A...528A..30C}, we reconstruct a sample of 140 galaxies with HCN 1-0, HCO$^+$ 1-0 and HNC 1-0 observations, containing 125, 124, 85 detections, respectively. 

\subsection{The  luminosities of dense gas tracers v.s.  infrared luminosities}

The correlations between  $L\rm_{IR}$ and $L'\rm_{dense~gas}$  (HCN 1-0, HCO$^+$ 1-0, and HNC 1-0, and CS 3-2) are presented in  Figure \ref{fig:correlation}, with the data listed in Table \ref{tab:luminosity}, including our data with new observations as well as  HCN 1-0, HCO$^+$ 1-0 and HNC 1-0 detections in local galaxies from literature \citep{2011A&A...528A..30C, 2015ApJ...814...39P}.  We adopt the Markov Chain Monte Carlo (MCMC) method using package emcee \citep{2013PASP..125..306F} to account for measurement uncertainties.
Only  detections are adopted to fit the relationship, while the upper limits are just  plotted in the Figures. The $L\rm_{IR}$ of sample galaxies span 7 orders of magnitude from a few 10$^5$L$_{\odot}$ to several times of 10$^{12}$L$_{\odot}$.  However, dense gas tracers are not detected in the low infrared luminosity galaxies, including NGC 6822, NGC 2403, and NGC 4605. Thus, the lowest infrared luminosity  with detection of dense gas tracers (HCN and HCO$^+$ 1-0) is  NGC 4605 with $log(L\rm_{IR})=7.82$,  while it is  $log(L\rm_{IR})=7.5$ in NGC 2976 with  CS 3-2 detection. Note that due to different beam sizes, there are large difference of  $log(L\rm_{IR})$  for regions corresponding to HCN 1-0 and CS 3-2 (see Table  \ref{tab:luminosity}), such as that in NGC 2976. The overall spanning of  $L\rm_{IR}$ is about 5 orders of magnitude  with detections of dense gas tracers.  
In the following, the fitted slope and uncertainties are the median and one standard deviation of the resulting MCMC chain:

\begin{equation}
\label{eq:HCN}
Log(L_{\rm IR})=0.97(\pm0.01)Log(L_{\rm HCN})+3.35(\pm0.09) 
\end{equation}

\begin{equation}
\label{eq:HCO}
Log(L_{\rm IR})=0.95(\pm0.01)Log(L_{\rm HCO^+})+3.59(\pm0.09) 
\end{equation}

\begin{equation}
\label{eq:HNC}
Log(L_{\rm IR})=0.97(\pm0.02)Log(L_{\rm HNC})+3.75(\pm0.11) 
\end{equation}

\begin{equation}
\label{eq:CS}
Log(L_{\rm IR})=1.07(\pm0.03)Log(L_{\rm CS})+2.95(\pm0.19) 
\end{equation}

The fitting results are  shown as the black solid line in Figure \ref{fig:correlation}, with a Spearman rank  correlation coefficient of 0.91, 0.94, 0.91 and 0.83, for HCN 1-0, HCO$^+$ 1-0, HNC 1-0, and CS 3-2, respectively. As shown with  blue dotted line in this figure,  only our IRAM data were used to fit the relationship, which gave a similar slope of 1.03$\pm$0.02, 0.99$\pm$0.02 and  1.01$\pm$0.02 for HCN 1-0, HCO$^+$ 1-0 and HNC 1-0, yielding a Spearman rank correlation coefficient  ($r_s$) of 0.83, 0.85 and 0.86, respectively, with  worse  Spearman rank correlation coefficient  than the whole sample, which  should mainly due to less sources in only our data than the whole sample. On the other hand, no clear difference of  correlation coefficient,  can be found among the  four tracers, which is from a Spearman rank correlation analysis.

\section{Discussion}

\subsection{Dense gas star formation law derived with different dense gas  tracers}

Dense molecular gas and star formation had been studied from the relation between  HCN 1-0 and infrared luminosities \citep{1992ApJ...387L..55S,2004ApJ...606..271G}, with a tight linear correlation.  However, unlike low$-J$ CO lines as  total molecular gas tracers,  there are many choices of dense gas tracers with transitions of   high dipole moment  molecules, such as HCN, HCO$^+$, HNC, CS, CN, and HC$_3$N.  Even though there are  dense gas tracers other than HCN 1-0  which also provide linear correlation in the past years: CS 5-4 \citep{2011MNRAS.416L..21W}, HCN 4-3, CS 7-6 and HCO$^+$ 4-3 \citep{2014ApJ...784L..31Z}, and HCN 3-2  \citep{2020PASJ...72...41L},  sub-linear relation  with slope of 0.79$\pm$0.09 for HCN 3-2  \citep{2008ApJ...681L..73B} and super-linear relation with HCN 1-0  \citep{2008A&A...479..703G,2012A&A...539A...8G} were also reported.  However, the sub-linear  slope of  $L\rm_{IR}$ and $L'\rm_{HCN_{3-2}}$ reported in \cite{2008ApJ...681L..73B} is mainly caused by  over-estimation of $L\rm_{IR}$  in galaxies with HCN 3-2 emission, because the beam size  of SMT for HCN 3-2 observations is smaller than the sizes of nearby galaxies. After correcting the infrared emission to the same as observed  HCN 3-2, a linear correlation was derived \citep{2020PASJ...72...41L}. On the other hand,  far-IR luminosities instead of IR luminosities are used in \cite{2008A&A...479..703G} and \cite{2012A&A...539A...8G}, which may be the main reason of super linear correlation since linear correlation was obtained with a similar sample by \cite{2015ApJ...814...39P}.

\begin{figure*} 
    \centering
  \includegraphics[height=2.5in,width=3.2in]{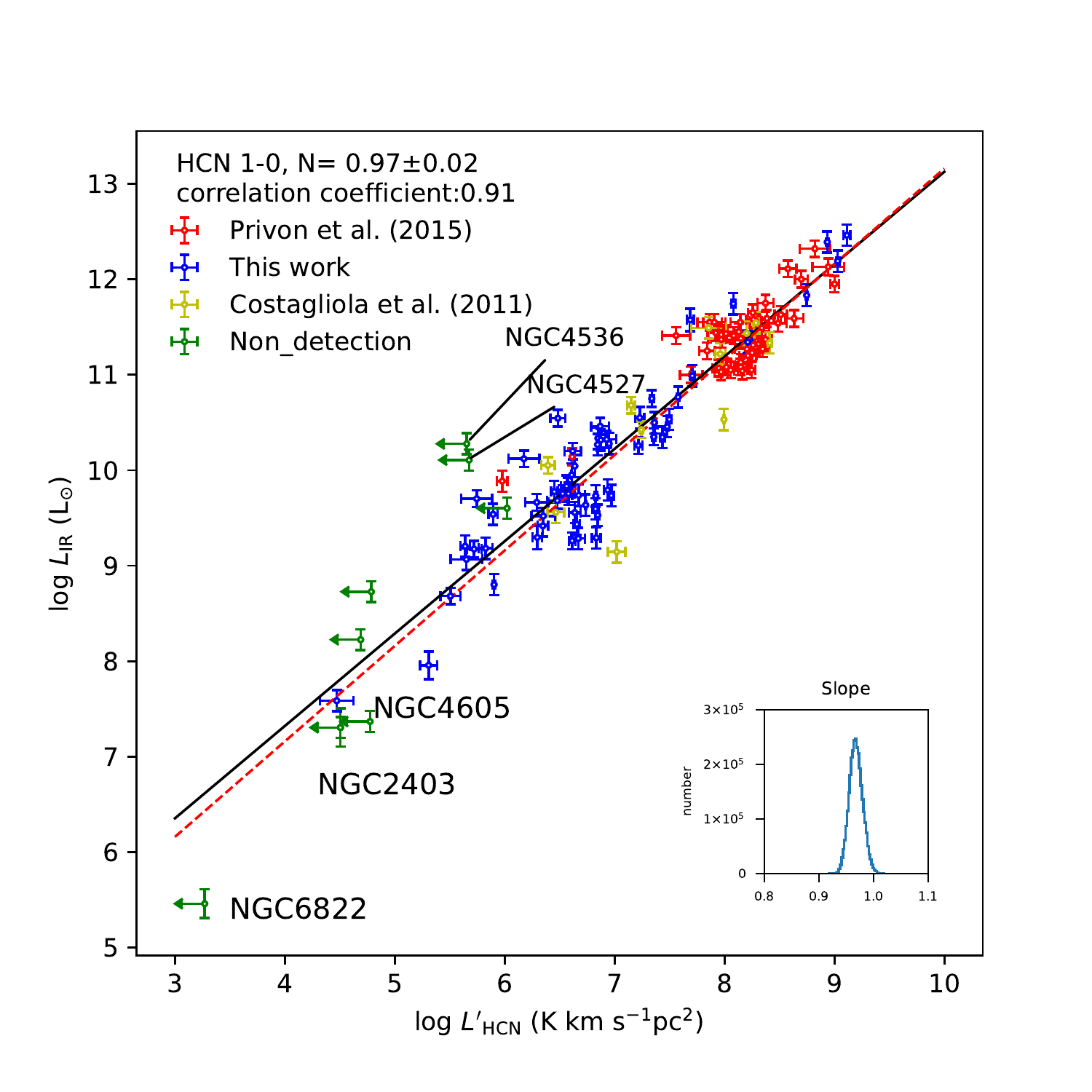}
   \includegraphics[height=2.5in,width=3.2in]{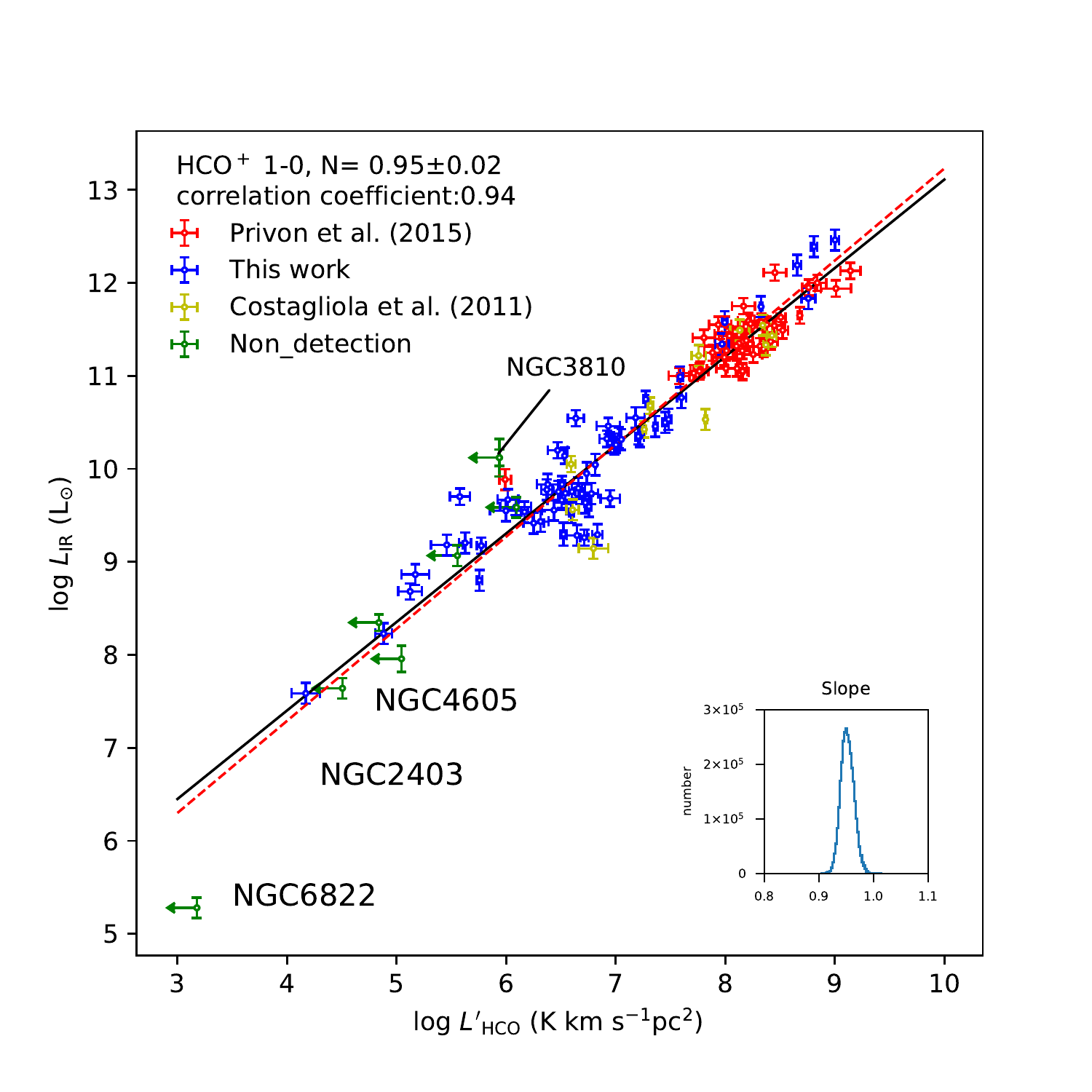}
     \includegraphics[height=2.5in,width=3.2in]{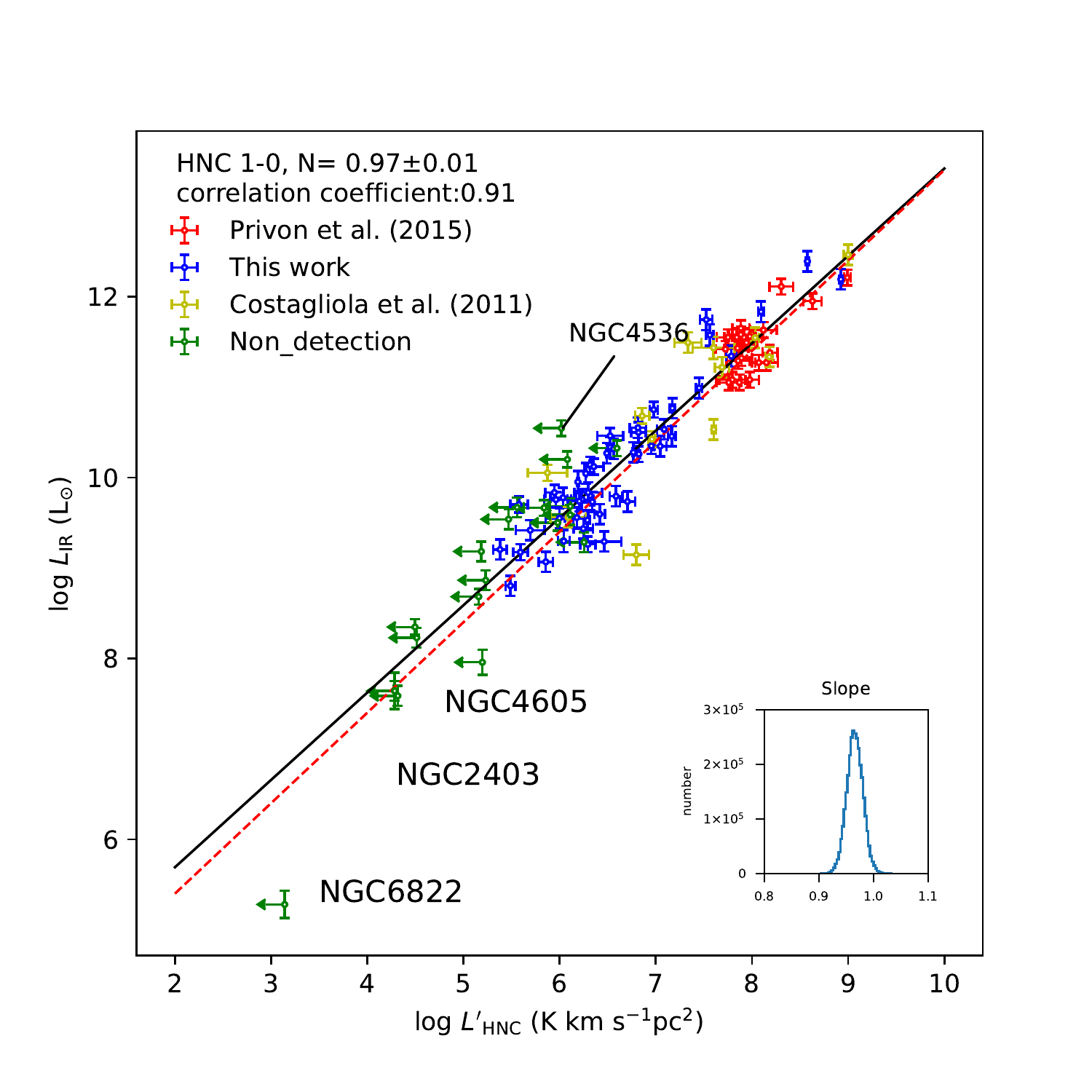}
       \includegraphics[height=2.5in,width=3.2in]{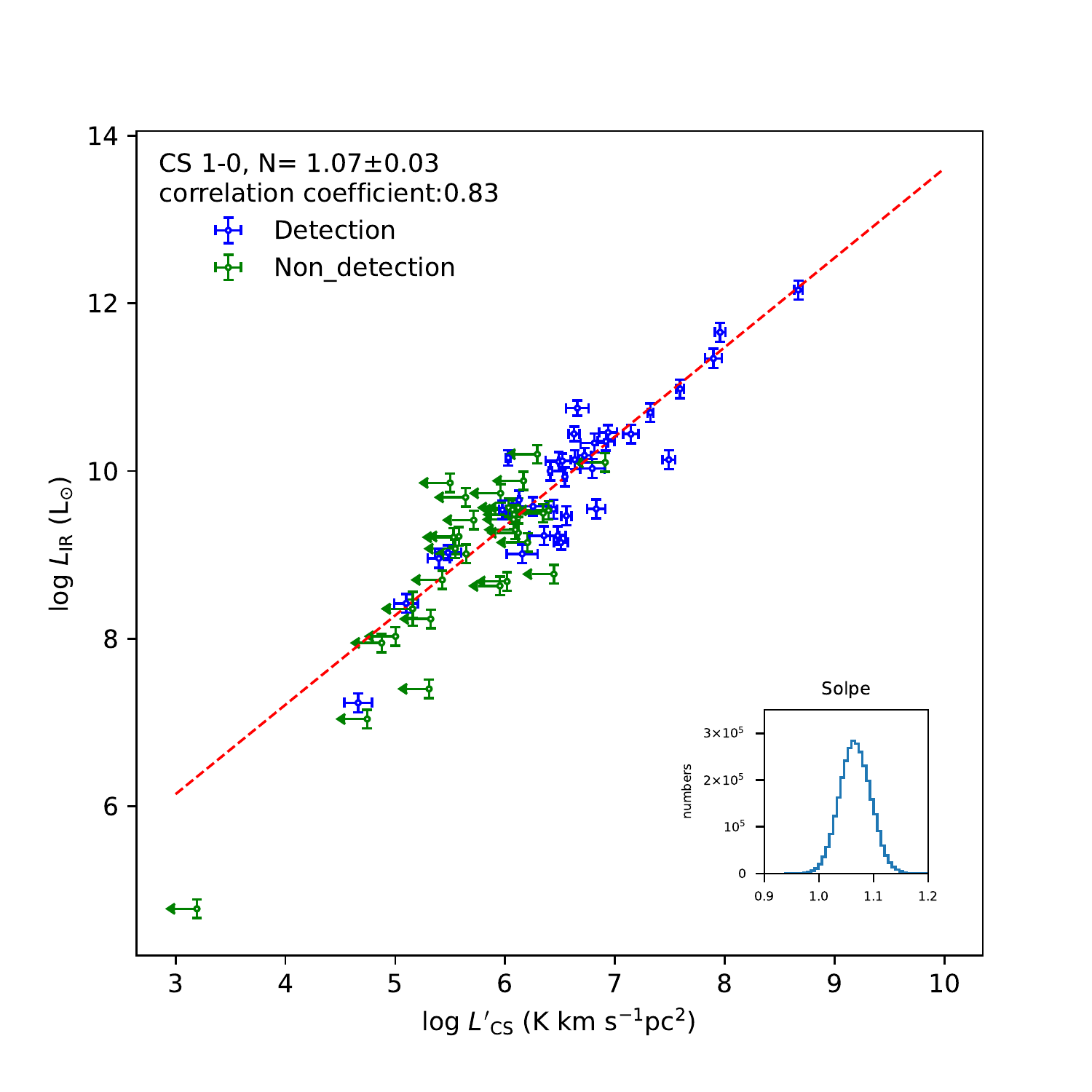}
      \caption{The relations between the gas luminosity log(L$^{'}\rm_{dense~gas}$) and the IR luminosity log(L$\rm_{IR}$). The circles are the sources with detected dense gas tracers from our data in blue, and  literature in red \citep{2015ApJ...814...39P} and yellow \citep{2011A&A...528A..30C}, respectively. The points in green  are  sources with non-detected dense gas tracers from our data and the literature data. The upper limits   are not adopted in the fitting. The inset panels show the probability density distribution of the slopes derived from the Bayesian fitting. Upper left: HCN 1-0, Upper right: HCO$^+$  1-0, Lower left: HNC 1-0, Lower right: CS 3-2.  In these four panels, the red solid lines indicate the best-fit relations of Equations (4)-(6) respectively, using the all data. While, the blue dotted lines show the relation using  our IRAM data alone.}
    \label{fig:correlation}
\end{figure*}

As shown in Figure \ref{fig:correlation}, the relations between the line luminosities  of HCN 1-0, HCO$^+$ 1-0, HNC 1-0, and  CS 3-2 and infrared luminosity are all close to linear correlation. All the data, including our observations and that from \cite{2015ApJ...814...39P} and \cite{2011A&A...528A..30C}, used in Figure \ref{fig:correlation}, are observed with IRAM 30m, which provides the same beam size for each line to avoid the problem of beam matching for different observations.  The  high Spearman rank  correlation coefficients between infrared luminosities and luminosities of  HCN 1-0, HCO$^+$ 1-0,  HNC 1-0 and, CS 3-2 with slopes close to unity indicate that even though there might be several times uncertainties of the conversion factor  from line luminosity of dense gas tracers to dense molecular gas mass in individual  galaxy,  no clear systematic bias of such conversion factor for different tracers.  The star formation law obtained with different  dense gas  tracers  is consistent with each other. Note that not only the detections follow the good correlation in Figure \ref{fig:correlation}, but also the upper limits  agree well with the relation, with infrared luminosities  spanning  7 orders of magnitude.  
  However, since the normally used dense gas tracers are optically thick, the self absorption of these lines, as well as the abundance issue due to different chemical conditions,  such as radiation field and shocks,   can cause large uncertainties of estimating dense gas mass from only one  line. Such uncertainties may be the main reason of the scatter in the relation plotted in Figure \ref{fig:correlation}. 

\subsection{Possible reason for sources with deviation from the correlation and note for several  individual sources}

Even though  the relations between  luminosities of dense gas tracers and infrared luminosities  are pretty tight with high Spearman rank  correlation coefficients for HCN 1-0, HCO$^+$ 1-0, HNC 1-0, and CS 3-2, there are large  scatters up to about 10 times for some sources (see Figure \ref{fig:correlation}).  Such deviation may be real for those sources because of  low or high star formation efficiency  for dense gas, or  the uncertainties of converting dense gas mass from line luminosity of dense gas  tracers, since the observational parameters are only  line fluxes, which can be used to obtain line luminosities with known distances.  
For the conversion  factor from line luminosity to dense gas mass, several properties can cause the uncertainty, including optical depth, excitation conditions (density and temperature), and abundance of  dense gas tracers caused by metallicity and other properties such as PDRs.  However, since  star formation rates, traced by infrared luminosities or hydrogen recombination lines, such as H$\alpha$,  is counting for  massive young stars, while dense gas traced by those dense gas tracers is counting for the gas that will form stars in the near future, the correlation can break  if there are large differences of  star formation evolutions in different  galaxies.   Another effect that can overestimate  star formation rate from infrared luminosity is AGN contribution, especially  in galaxies with low star formation  activities.

There are several sources, which are not detected in  one or more lines,  should be noted for deviation from the correlation in Figure \ref{fig:correlation}.  No detections were found  for all four lines in  NGC 2403, NGC 4605 and NGC 6822. Even though they are in the regions away from the fitting lines,  they are still consistent with the  correlation, since they are upper limits of line emission. If the lines can be detected with more sensitive  observations, it should be closer to the fited lines or even at the other side of the lines. Based on the current results, the real deviations can be seen for  NGC 4527, NGC 4536, NGC 3810, and NGC 5907.   Especially,  only HCO$^+$ 1-0 is detected in NGC 4536 at 4.5 $\sigma$ level, while other three lines remain non-detected.  If the conversion  factor of dense gas mass from dense gas tracer luminosity in these sources are similar to that of other galaxies, it will imply that there are less dense gas than expected.  In NGC 4536, the metallicity is within normal range and with strong CO 1-0 (525.67$\pm$4.49 Jy km s$^{-1}$) and $^{13}$CO 1-0  (54.51$\pm$3.98 Jy km s$^{-1}$)  emission  \citep{2017ApJ...847...33C}. While even HCO$^+$ 1-0 detected in NGC 4536 is only 0.23$\pm$0.05 K km s$^{-1}$, which corresponds to about 1.2 Jy km s$^{-1}$. The dense gas fraction is really low, based on the line ratio of HCN and HCO$^+$ 1-0 to CO and    $^{13}$CO 1-0.  For comparison, the line ratio of CO/HCN 1-0 in starburst galaxies are normally around 10 \citep{2002A&A...381..783A}, much lower than than in NGC 4536. No AGN activity is found in NGC 4536. Thus, we suggest that the most possible reason  for  the lack of dense gas in these galaxies, including NGC 4536, is that star formation activities are at late evolutionary stages with strong feedback to the molecular gas, which causes the  low dense gas fraction.

\subsection{Star formation efficiency v.s. infrared luminosity revealed by different dense gas tracers}

Star formation efficiency (SFE) of dense gas, SFR/$M_{\rm dense}$, can be derived with $L_{\rm IR}/L_{\rm dense}$. With  $L_{\rm FIR}/L_{\rm HCN}$ ratio for galaxies with $L_{\rm FIR}$ between 10$^{10}L_\odot$ to several times of 10$^{13}L_\odot$, \cite{2012A&A...539A...8G} claimed that SFE increased with the increment of SFR ($L_{\rm FIR}$), with large scatter.  However, no clear trends of $L'_{\rm HCN}/L_{\rm IR}$ and  $L_{\rm IR}$,  or $L_{\rm HCO^{+}}/L_{\rm IR}$ and  $L_{\rm IR}$  were  found in \cite{2015ApJ...814...39P}.  

\begin{figure*} 
    \centering
    \includegraphics[height=2.5in,width=3.2in]{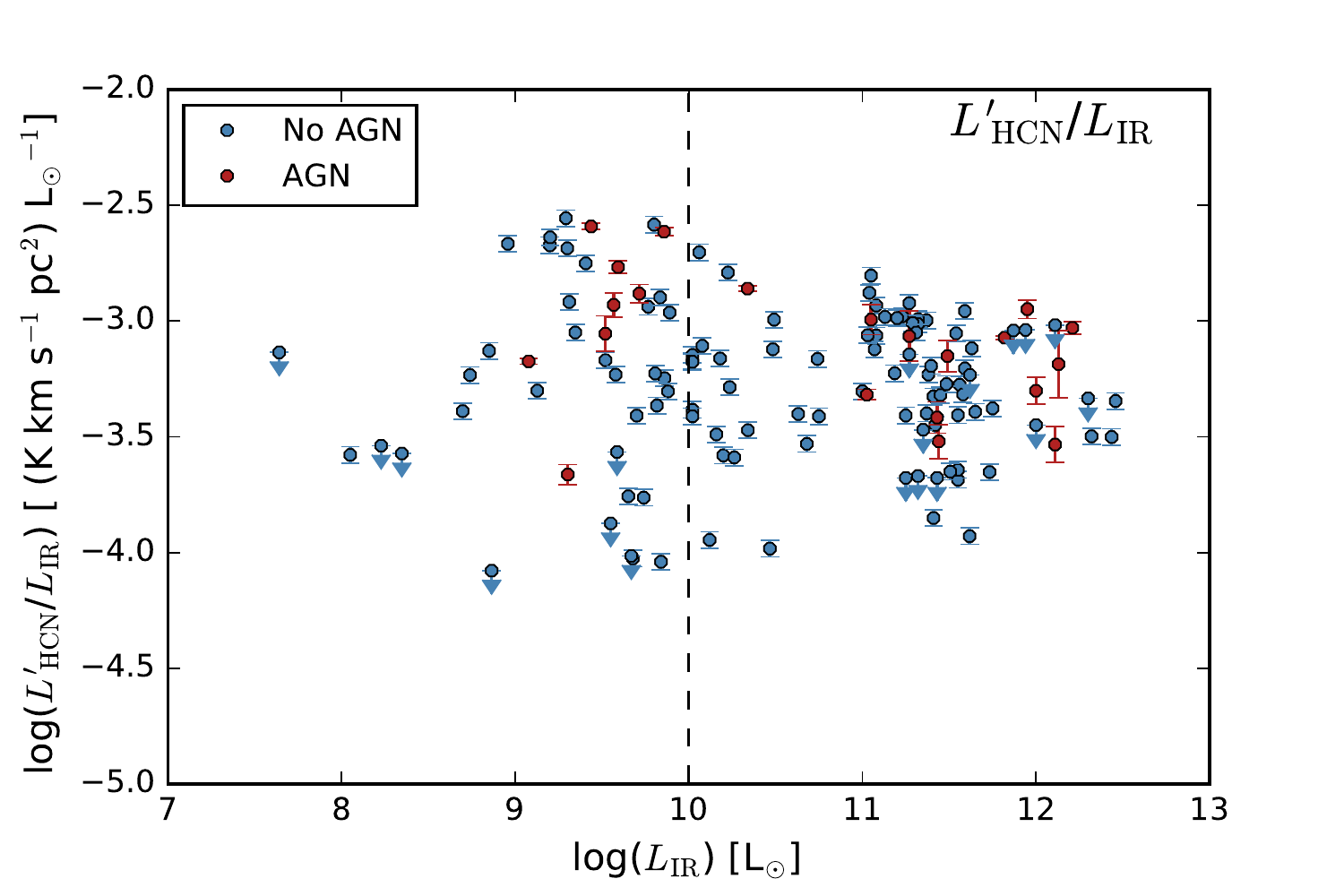}
     \includegraphics[height=2.5in,width=3.2in]{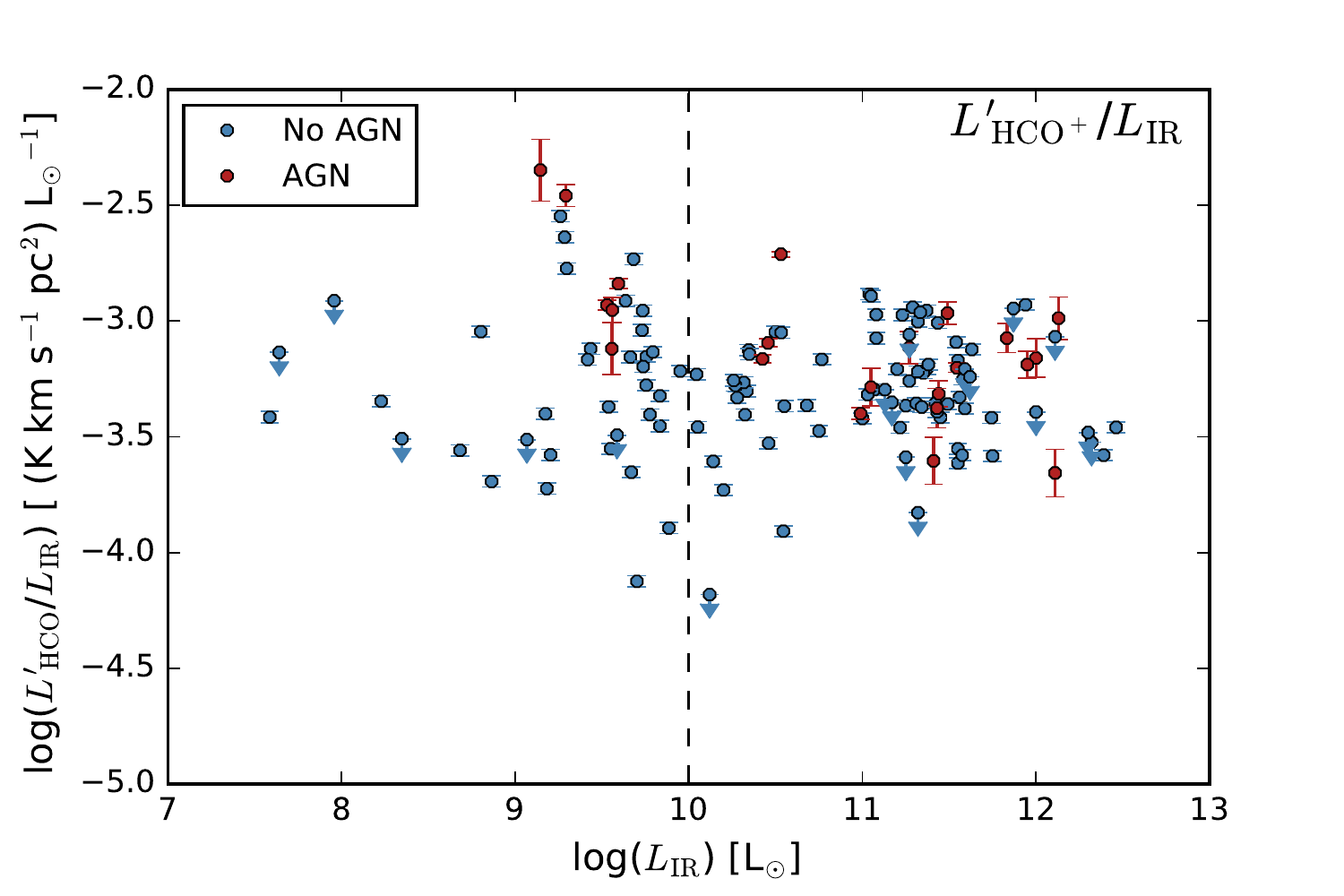}
      \includegraphics[height=2.5in,width=3.2in]{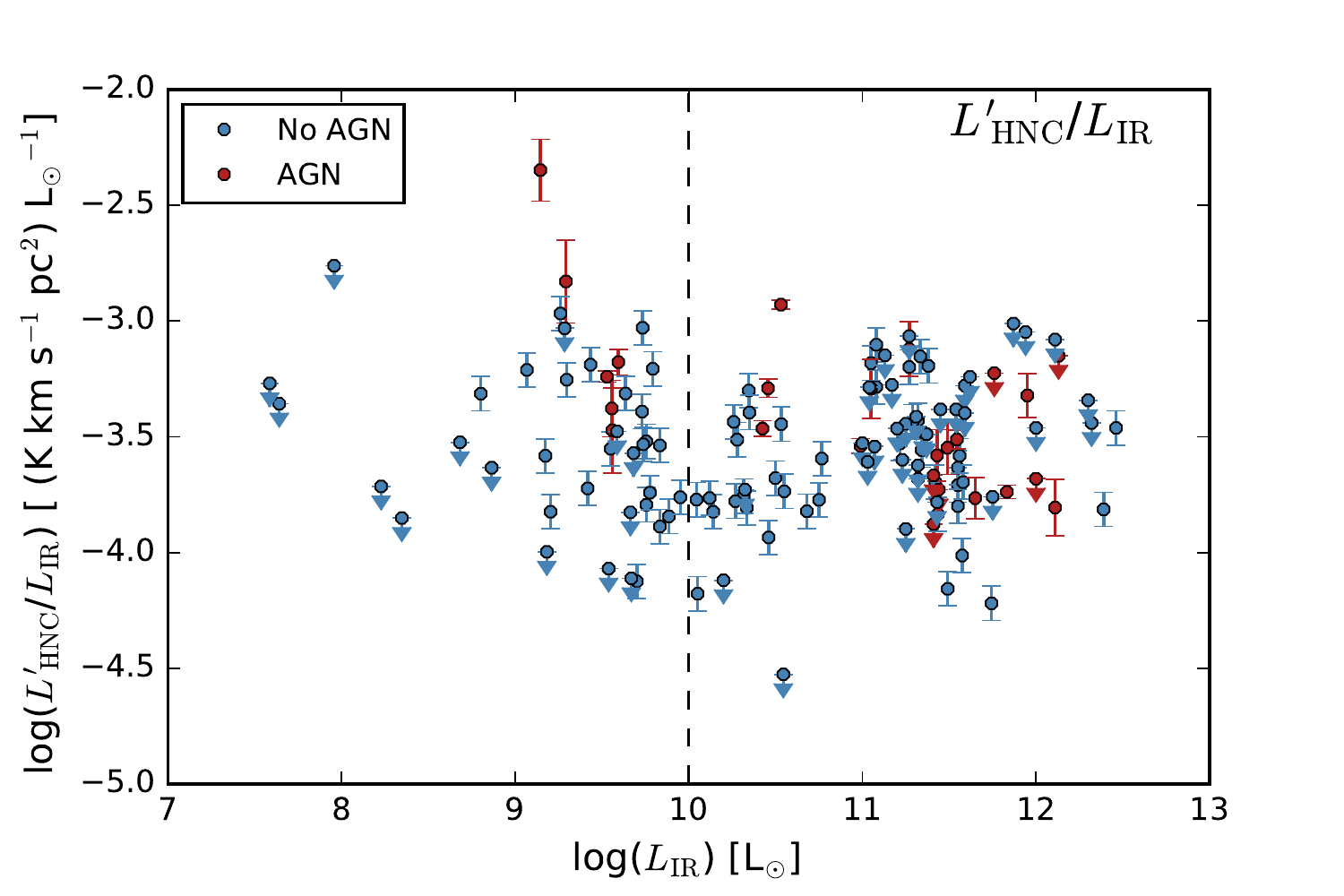}
    \caption{The relation between star formation efficiency derived with different dense gas tracers and infrared luminosities. Upper left: L$'\rm_{HCN}$/L$\rm_{IR}$ vs. $L\rm_{IR}$. Upper right: L$'\rm_{HCO^+}$/L$\rm_{IR}$ vs.$L\rm_{IR}$. Lower: L$'\rm_{HNC}$/L$\rm_{IR}$ vs.$L\rm_{IR}$. These sources are from Table \ref{tab:luminosity}.  The sources marked as ``AGN" are identified with AGN activity from literature, while sources marked as ``No AGN" are without known AGN activity in the literature. However, some of   ``No AGN" sources  may still be with AGN activity.}
    \label{fig:SFE}
\end{figure*}

Figure \ref{fig:SFE} shows the $L'_{\rm dense}$/$L_{\rm IR}$ versus $L_{\rm IR}$, including our new data and literature data from \cite{2015ApJ...814...39P} and \cite{2011A&A...528A..30C}. No trend of $L'_{\rm dense}$/$L_{\rm IR}$ versus $L_{\rm IR}$ can be found for  HCN 1-0, HCO$^+$ 1-0, and HNC 1-0, with $L_{\rm IR}$ spanning range larger than that discussed in \cite{2012A&A...539A...8G}  and \cite{2015ApJ...814...39P}.   On the other hand, $L'_{\rm dense}$/$L_{\rm IR}$ can vary more than ten times for different galaxies with similar $L_{\rm IR}$.  In figures \ref{fig:SFE}, \ref{fig:LIR_ratio}, and \ref{fig:HCN_HCO_HNC}, the sources marked as ``AGN" are identified with AGN activity from NED\footnote{The NASA/IPAC Extragalactic Database (NED) is funded by the National Aeronautics and Space Administration and operated by the California Institute of Technology.}, while sources marked as ``No AGN" are without known AGN activity  also from NED. However, some of   ``No AGN" sources  may still be with AGN activity.

 Sources with known AGN activities do not show clear difference when compared with those without known AGN activities.  Thus, the dense  gas depletion time to form stars from dense molecular gas, does not relate to SFR traced by infrared luminosity.  No bias can be found for different dense gas tracers (HCN 1-0, HCO$^+$ 1-0 and HNC 1-0), or with or without known AGN activities. We would like to suggest that  star formation from dense molecular gas follows the similar law in different galaxies: low or high SFR, with or without known AGN. However, the dense gas fraction: the ratio of dense gas to total molecular gas,  and  molecular gas to atomic gas ratio, which are important to affect star formation from gas, still need to be studied  for  understanding star formation efficiency in different types of galaxies.   Metallicities, merging history, AGN feedback, and other properties, can affect star formation in galaxies.  The formation from atomic gas to molecular gas, the collapse of molecular clouds to dense cores, should be understood in further studies, which may also be compared with gas to star ratios in different galaxies.

\subsection{Line ratios between different dense gas tracers in individual galaxies}

HCN/HCO$^+$ line ratio had been used to  distinguish between AGNs and starburst signatures in galactic centers \citep{2001ASPC..249..672K,2004AJ....128.2037I,2007AJ....134.2366I,2008ApJ...677..262K}, while HCN/HNC ratio together with  HCN/CN ratio had beed used to discuss starburst evolution in galaxies \citep{2002A&A...381..783A}.  With HCN/HCO$^+$ 4-3 ratio and  CS 7-6 ratio in a sample of galaxies with AGN or without known AGN, \cite{2013PASJ...65..100I,2016ApJ...818...42I}  and \cite{2008ApJ...677..262K} suggested  that HCN can be enhanced in AGNs and HCN/HCO$^+$ line ratio can be used to diagnose AGN and starburst contribution in galaxies.  Lower average HCN/HCO$^+$ 1-0 line ratio in LIRGS and starbursts than that in  AGNs was found in a limited sample \citep{2011A&A...528A..30C}, which was consistent with 
 the weak trend found in \cite{2001ASPC..249..672K} and \cite{2007AJ....134.2366I}.

\begin{figure*} 
    \centering
      \includegraphics[height=2.5in,width=3.2in]{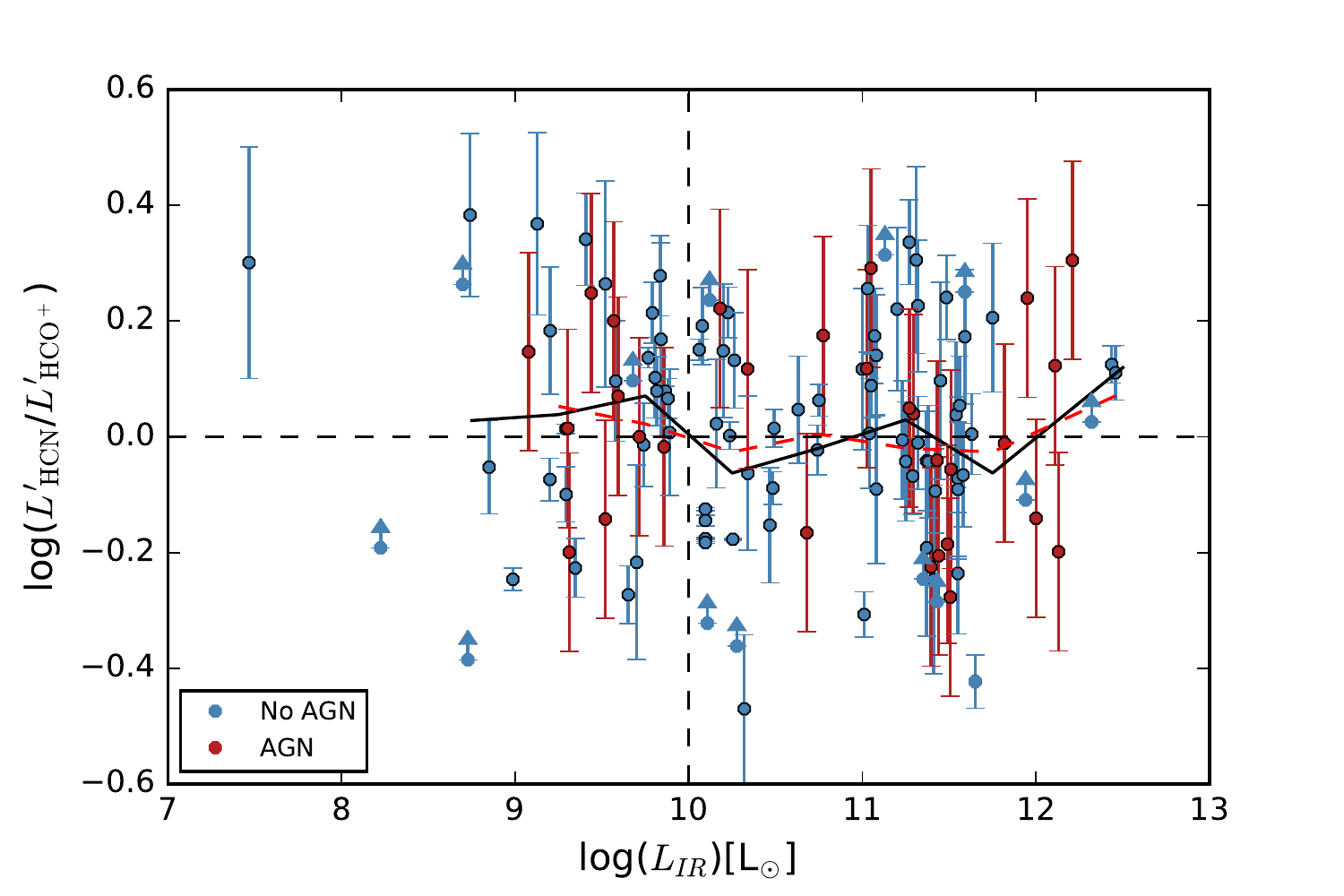}
     \includegraphics[height=2.5in,width=3.2in]{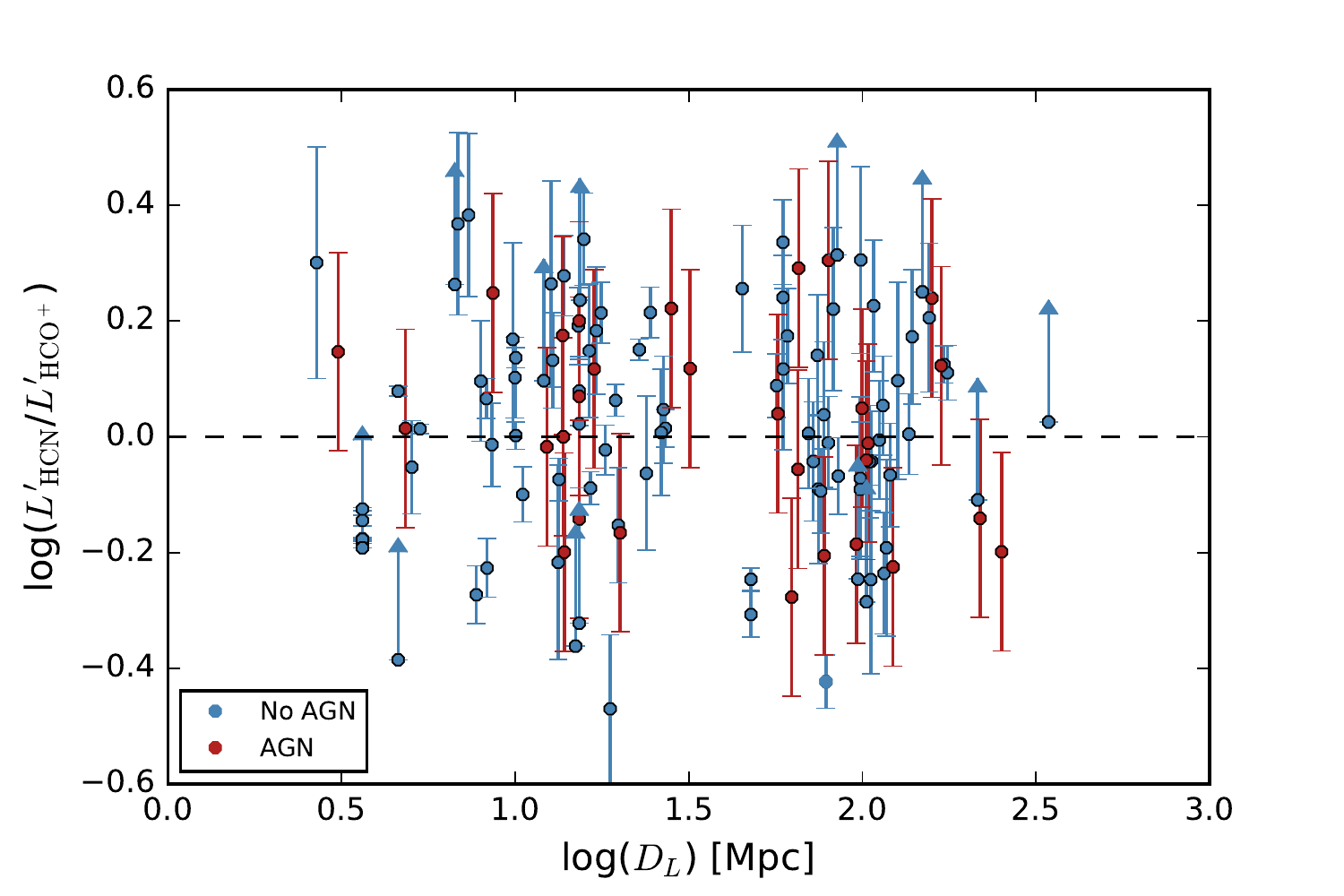}
    \includegraphics[height=2.5in,width=3.2in]{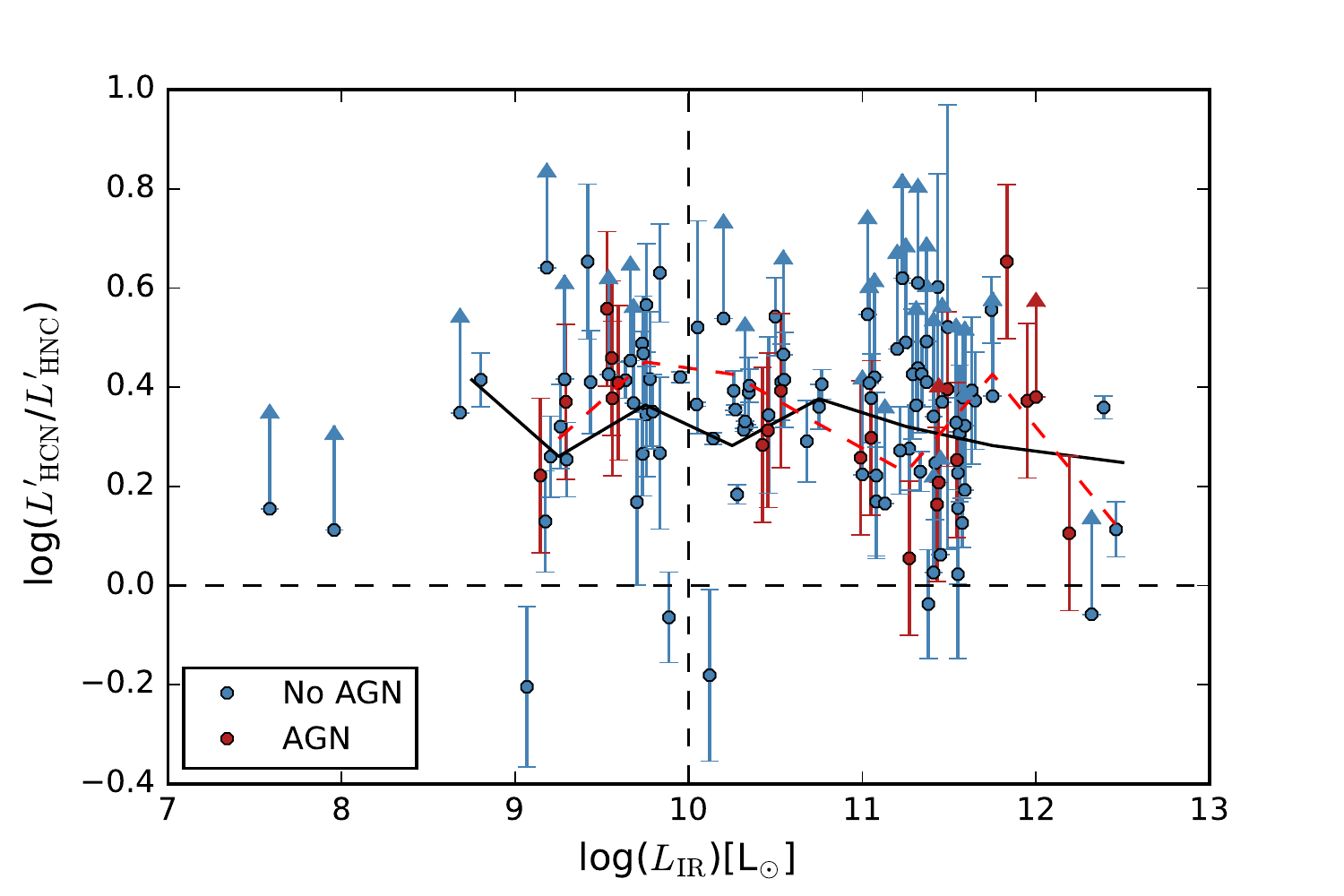}
     \includegraphics[height=2.5in,width=3.2in]{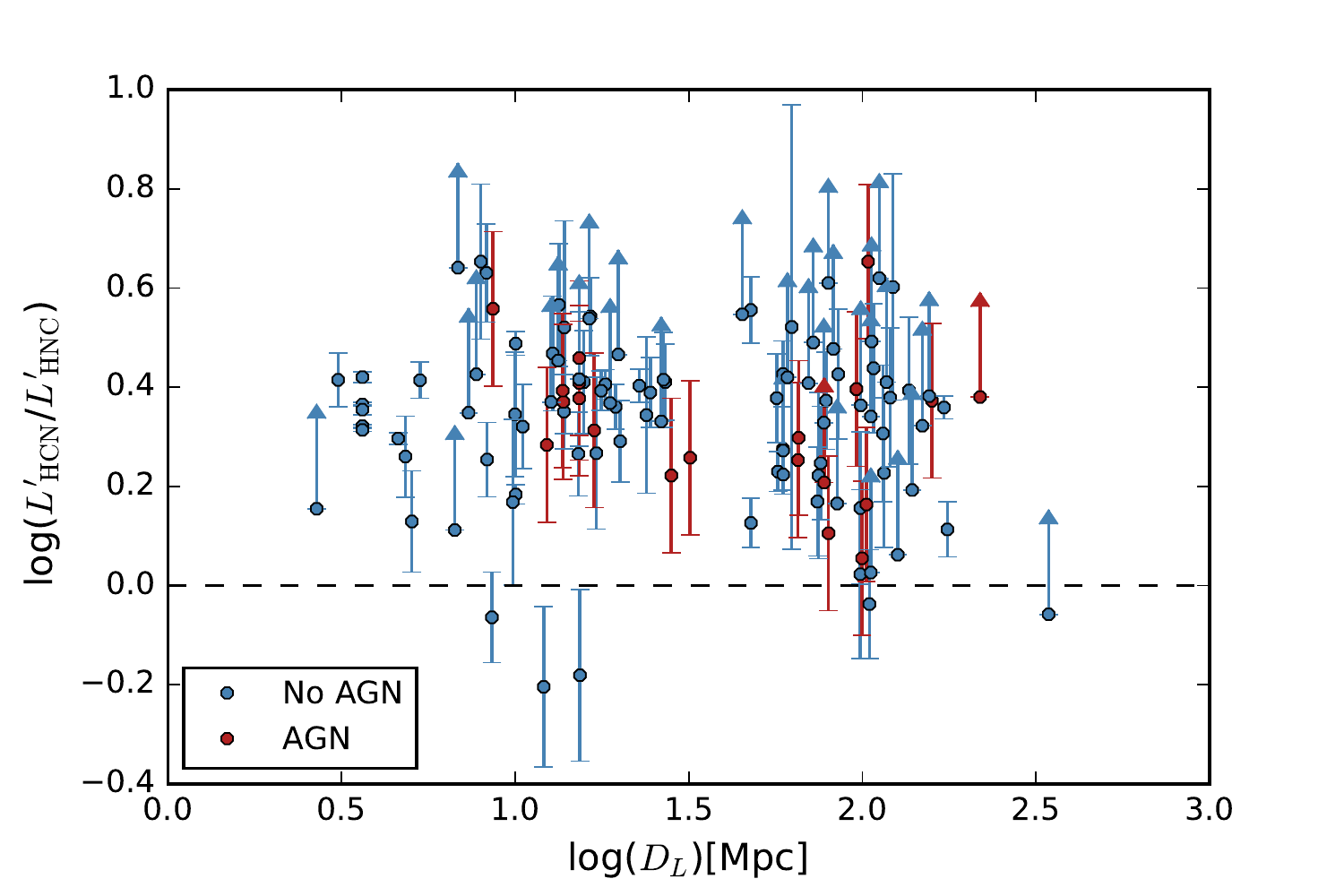}
    \caption{HCN/HCO$^+$ 1-0 line ratios as a function of IR luminosity (upper left) and luminosity distance (upper right), as well as HCN/HNC 1-0 ratio as a function of IR luminosity (lower left ) and luminosity distance (lower right). In these four panels, the black solid lines indicate mean luminosity ratios of dense gas tracers for the sources without known  AGN activity. While, the red dotted lines show the result from the source with AGN activity.  The black dotted lines imply that  ratios equal one.}
    \label{fig:LIR_ratio}
\end{figure*}

However, with limited sample of galaxies in \cite{2008ApJ...677..262K},  \cite{2011A&A...528A..30C} and \cite{2013PASJ...65..100I,2016ApJ...818...42I}, it is hard to conclude if such line ratios can be used to diagnose AGN and/or  starburst  in galaxies, since some exceptions of HCN/HCO$^+$ line ratio had been found in literature: a sample of (U)LIRGs \citep{2015ApJ...814...39P}, and NGC 4258 with AGN \citep{2019MNRAS.482.4763L}.  With the sample including our observations and that from \cite{2011A&A...528A..30C}  and \cite{2015ApJ...814...39P}, we complied a sample with more than 100 galaxies, with detection of a least  two lines of HCN, HCO$^+$, and HNC 1-0.  The plots of line ratios of HCN/HCO$^+$ 1-0 and  HCN/HNC 1-0 versus  infrared luminosities and luminosity distances are presented in Figure \ref{fig:LIR_ratio}.  HCN/HCO$^+$ 1-0  ratio varies from about  0.38 in IC 1623 \citep{2015ApJ...814...39P} to 2.2  in NGC 4535 (see Table \ref{tab:intensity}), while HCN/HNC 1-0 ratio varies from 0.86 in NGC 891 to 4.3 in NGC 2903 (see Table \ref{tab:intensity}).  No clear trend was found for the line  ratios and  infrared luminosities, which is consistent with the results for HCN/HCO$^+$ 1-0 ratio in \cite{2015ApJ...814...39P}.  With HCN and HCO$^+$ 1-0 data  in a  large sample of galaxies    observed simultaneously   in \cite{2011A&A...528A..30C}, \cite{2015ApJ...814...39P} and our observations, which can avoid errors of flux ratio due to the uncertainties of  absolute flux calibration   and  pointing errors with different observations,   HCN/HCO$^+$ 1-0 ratios in these galaxies do not vary with different infrared luminosities,  however, with large scatters even with similar  infrared luminosities (see Figure \ref{fig:LIR_ratio}).  Thus, different   HCN/HCO$^+$ 1-0 ratios in galaxies should not be caused by star formation activities, as suggested by \cite{2006ApJ...640L.135G} for higher HCN/HCO$^+$ ratios in ULIRGs than galaxies with lower infrared luminosity.

 Red dashed  lines represent averaged values with $x-$axis bins for   ``AGN'' sources, while black lines for ``No AGN'' sources, are also presented in Figure \ref{fig:LIR_ratio}.   No clear difference between these two sub-groups:  ``AGN''  and  ``No AGN'', is found in Figure  \ref{fig:LIR_ratio}.   Abundance, opacity and excitation of HCN and HCO$^+$ molecules can affect the line ratio of HCN/HCO$^+$ 1-0. Even if starburst activities, which cause strong photon dominated regions (PDRs), or AGN activities, which cause strong X-ray dominated regions (XDRs), may enhance  HCN or HCO$^+$ abundances, it is hard to use line ratio of such  optically thick lines: HCN and HCO$^+$ 1-0, to derive the abundance ratio of HCN and HCO$^+$.  Such line ratio may not be a  good tracer of AGN activity had been pointed out by \cite{2011A&A...528A..30C} and \cite{2015ApJ...814...39P} with smaller sample than ours.   With the comparison with X-ray emission, \cite{2020ApJ...893..149P} had pointed out that HCN/HCO$^+$ line ratios can not be a reliable method  to find AGN, which is consistent with our results.

 As optically thin lines, H$^{13}$CN and H$^{13}$CO$^+$ 1-0, the isotopic lines of HCN and HCO$^+$, will be a better choice than  HCN and HCO$^+$ 1-0 to derive abundance ratio of HCN and HCO$^+$. However, it is hard to do large sample survey due to weak emission of these isotopic lines, which are normally several times to tens of times weaker than their main isotopic lines, as seen in  NGC 1068 \citep{2014ApJ...796...57W}, Arp 220 \citep{2016MNRAS.455.3986W}, Mrk 231 \citep{2016A&A...587A..15L,2020MNRAS.494.1095L}, M82  and NGC 3079 \citep{2020MNRAS.494.1095L}, and NGC 4418  \citep{2015A&A...582A..91C,2020MNRAS.494.1095L}. ALMA observations of  H$^{13}$CN and H$^{13}$CO$^+$ lines toward local galaxies in the future will help us to determine the enhancement of these molecules due to starburst or AGN activities.

  Neither  any trend of HCN/HNC 1-0 ratio along with infrared luminosity, nor clear difference between ``AGN'' and ``No AGN''  groups for  HCN/HNC 1-0 ratios versus infrared luminosities, can be found  (see Figure \ref{fig:LIR_ratio}).  We also notice that  the sources with low HCN/HNC 1-0 ratios in \cite{2002A&A...381..783A}:  1.0  in IC 694  and Mrk 231 may not be real.  With better sensitivity than that in  \cite{2002A&A...381..783A} and simultaneous observation, our data show that this ratio is 3.6$\pm$0.4 in IC 694 and 2.3$\pm$0.1 in Mrk 231 (see Table \ref{tab:intensity}).  Especially, the  simultaneous observation can avoid the uncertainties of pointing error to obtain the line ratio.  Similar to that of determining  possible enhancement of HCN or HCO$^+$ molecules due to starburst or AGN activities,  optically thin isotopic lines of HCN and HNC 1-0 should be used to derive HCN/HNC abundance ratio, even if this ratio can reflect starburst evolution with gas phase chemistry in molecular clouds. 
 
 HCN/HCO$^+$ 1-0 and  HCN/HNC 1-0 ratios versus luminosity distances are also presented in Figure \ref{fig:LIR_ratio}. Since the beam of IRAM 30m at 3mm band can cover only central region of nearby galaxies, while it can cover entire galaxy  for  distant sources. There is  no obvious  trend  of  HCN/HCO$^+$ 1-0 and  HCN/HNC 1-0  ratios with distance (see Figure  \ref{fig:LIR_ratio}). Thus, the covering area of one galaxy with IRAM 30m beam ($\sim28''$) is not a dominate factor for the line ratios, which  is consistent with that  in \cite{2015ApJ...814...39P}.

\begin{figure*} 
    \centering
   \includegraphics[height=2.5in,width=3.2in]{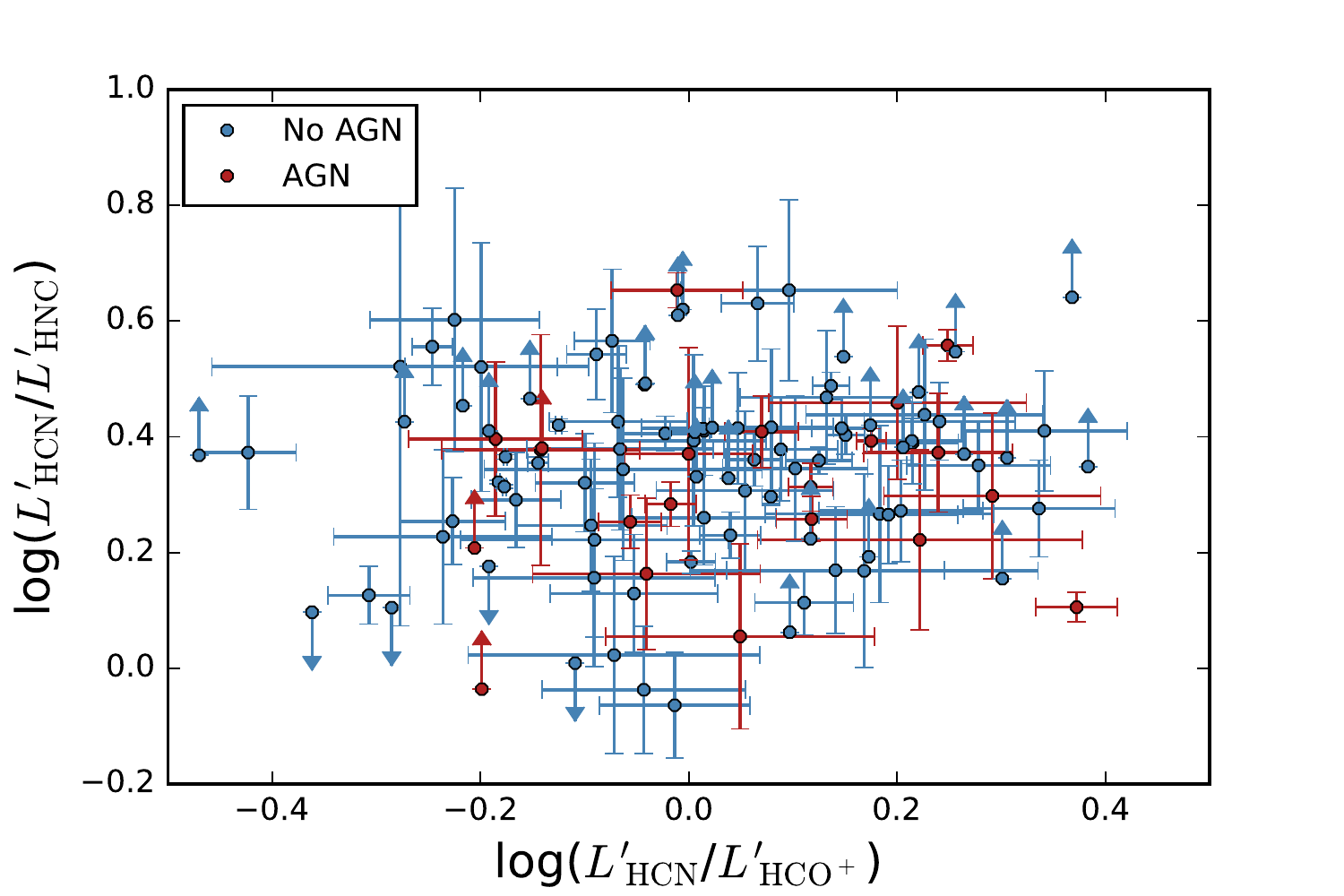}
   \includegraphics[height=2.5in,width=3.2in]{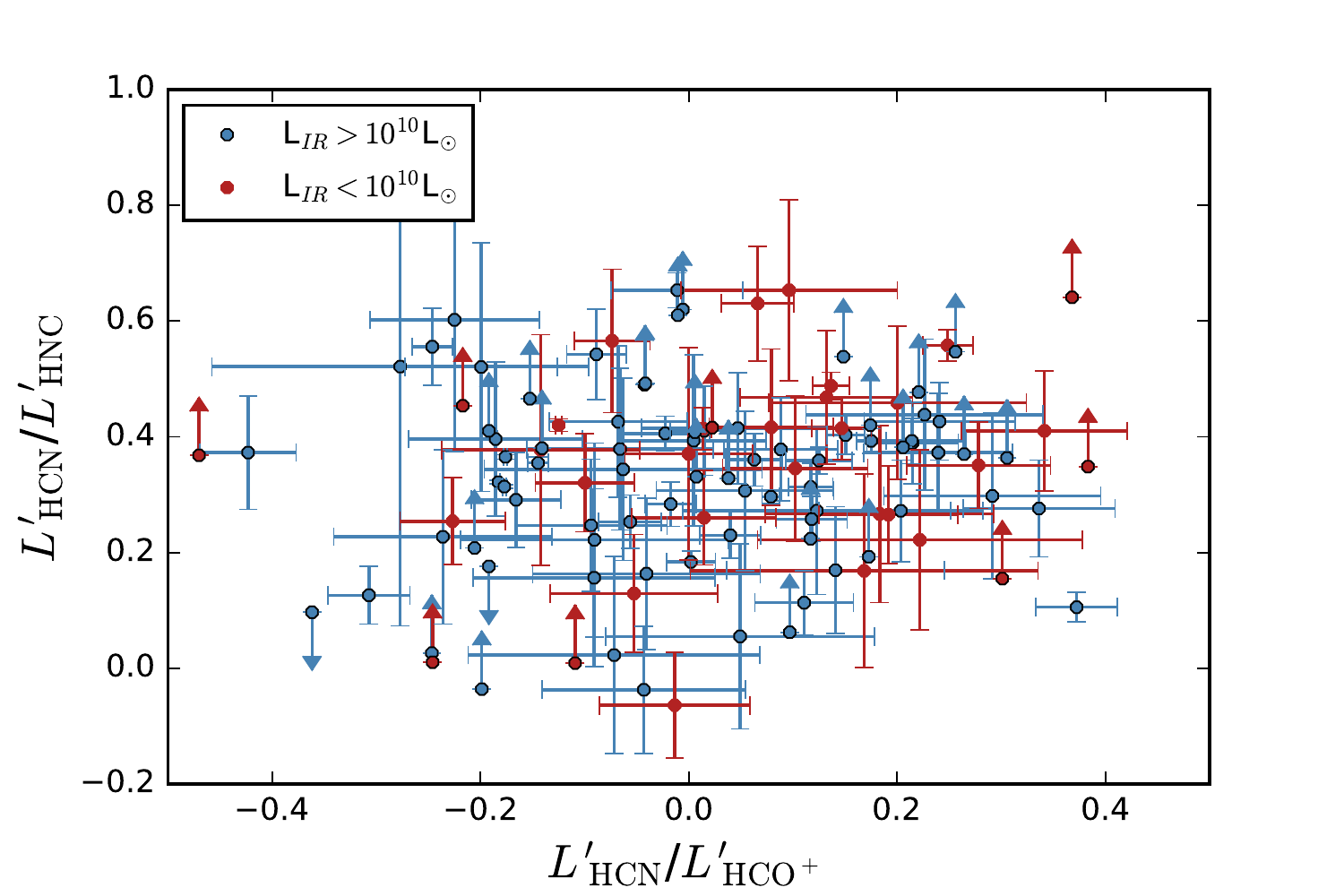}
     \caption{HCN/HCO$^+$ 1-0 ratios and  HCN/HNC 1-0 ratios in this sample. Left: ``AGN'' with red points and ``No AGN'' with blue points. Right: red points for sources with infrared luminosities less than 10$^{10}L_\odot$ and blue points for infrared luminosities greater than 10$^{10}L_\odot$.}
    \label{fig:HCN_HCO_HNC}
\end{figure*}

We also try to use the combination of  HCN/HCO$^+$ 1-0 and  HCN/HNC 1-0 ratio to determine the effect of  AGN and starburst activities.  Plots of HCN/HCO$^+$ 1-0 and  HCN/HNC 1-0 ratios  are presented in Figure \ref{fig:HCN_HCO_HNC}.  In the left of Figure \ref{fig:HCN_HCO_HNC}, the sample is separated into two sub-groups as ``AGN'' in red and ``No AGN'' in blue, while it is separated into two sub-groups with  $L_{\rm IR}>10^{10}L_\odot$ in blue and $L_{\rm IR}<10^{10}L_\odot$  in red at the right of Figure \ref{fig:HCN_HCO_HNC}.  It is impossible to separate   ``AGN''  and ``No AGN'', or starburst and normal star forming galaxies, with the combination of HCN/HCO$^+$ 1-0 and  HCN/HNC 1-0 ratios.

\section{summary and conclusion remarks }

With simultaneous  observations of  HCN 1-0, HCO$^+$ 1-0, HNC 1-0, and CS 3-2 toward 70 local galaxies with a wide infrared luminosity coverage from several 10$^5L_\odot$ to more than 10$^{12}L_\odot$, we detected these lines with detection rate of 89\%, 90\%, 73\%, and 51\%, respectively.  After combining data from literature  \citep{2015ApJ...814...39P,2011A&A...528A..30C}, we complied a sample of 140 galaxies  with simultaneous observations of HCN 1-0, HCO$^+$ 1-0, and HNC 1-0.  With comparison of  luminosities of dense gas  tracers and infrared luminosity,   line ratios, the main conclusions are listed below:

1.  The slopes of global relations between luminosities of dense gas tracers (HCN 1-0, HCO$^+$ 1-0, HNC 1-0, and CS 3-2) and   infrared luminosities are close to unity.  Even though there may be several times uncertainties of   estimating dense gas mass in individual galaxy  from one dense gas tracer,  these dense gas tracers give  similar results for dense gas star formation law.

2. No clear trend is found for the relation between infrared luminosities and star formation efficiencies  derived with HCN 1-0, HCO$^+$ 1-0 and  HNC 1-0, which means that dense gas depletion time to form stars from dense molecular gas does not relate to SFR. 

3. HCN/HCO$^+$ line ratio can not be a good tool to diagnose obscured  AGN  in galaxies. Even combining with HCN/HNC line ratio will not improve such diagnostic.

\section*{Acknowledgements}

This work is supported by  the National Natural Science Foundation
of China grant 11590783, and U1731237.  This study is based on observations
carried out under project number 066-19, 186-18 and 058-17  with the IRAM 30-m telescope. IRAM is
supported by INSU/CNRS (France), MPG (Germany) and IGN (Spain).  This research has made use of the NASA/IPAC Extragalactic Database, which is funded by the National Aeronautics and Space Administration and operated by the California Institute of Technology.

\section*{Data availability}
The original   data observed with IRAM 30 meter  can be accessed by IRAM archive system at https://www.iram-institute.org/EN/content-page-386-7-386-0-0-0.html. If anyone is interested in the  reduced  data presented in this paper, please contact  Junzhi Wang at jzwang@shao.ac.cn.



\clearpage

\onecolumn

\begin{center}

\begin{longtable}{llllllllllllllll}

\caption{Galaxies Observed with IRAM 30-m Telescope}\label{tab:observation parameters}\\
		
   	         \hline
	         \hline
	     
	         {Source Name} & {RA} & {DEC}  & {F$_{100\mu m}$}  &{cz}   &{Distance} & {Obs Month} &{t$_{\rm int}$} &{T$_{sys}$}  \\
                             & (J2000)  & (J2000)  &(Jy)       &   km s$^{-1}$     & (Mpc) &(YYYY MM) &(minutes) &(K)  \\
	         \hline
	          \hline
	\endfirsthead
\caption[]{(continued)}\\
\hline 
\hline

	         {Source Name} & {RA} & {DEC}  & {F$_{100\mu m}$}  &{cz}   &{Distance} & {Obs Month} &{t$_{\rm int}$} &{T$_{sys}$}  \\
                             & (J2000)  & (J2000)  &(Jy)       &    km s$^{-1}$   & (Mpc) &(YYYY MM) &(minutes) &(K)  \\
	         \hline
	          \hline

    \endhead	          
\hline
\endfoot
%

NGC 891         &02:22:33.5          &  +42:21:18          &  172.23         &  533             &  8.57       & 2019 Aug     &80  &103\\   
IC 342               &03:46:49.4          &  +68:05:49          &  391.66          &   31              &  4.60       &  2019 Aug     &20  &100\\   
UGC 02855       &03:48:19.4          &  +70:08:03          &   89.18          & 1184              & 19.46      & 2019 Aug     &100  &108\\  
NGC 1569         &04:30:49.5          &  +64:51:01          &   55.29          &  -97                &  4.60       & 2019 Aug     &102  &126\\    
NGC 2146         &06:18:39.8          &  +78:21:25          &  194.05          &  885                 & 16.47       &  2019 Aug     &61  &122\\ 
NGC 2403         &07:36:51.3          &  +65:36:30          &   99.13          &  161                  &  3.22      & 2019 Aug     &81  &101\\ 
NGC 2683         &08:52:40.1          &  +33:25:23          &   30.68          &  425                &  7.34      &  2019 Aug     &204  &122\\  
NGC 2903         &09:32:10.5         &  +21:30:05          &  130.43         &  566          &  8.26        &2019 Jul      &37   &99 \\
NGC 2976         &09:47:14.5          &  +67:55:10          &   33.43          &   11                   &  2.68        &  2019 Aug     &86  &106\\  
NGC 3031         &09:55:33.6          &  +69:03:56          &  174.02          &  -34                  &  3.63        &  2019 Aug     &88  &107\\ 
M82\_1               &09:55:48.0             &  +69:40:40             &1373.69          &187                 &3.63            &2019 Feb       &115  &83\\
M82\_2        &09:55:50.0            &  +69:40:43             &1373.69          &187                 &3.63            &2019 Feb       &116  &94\\
M82\_3(center)                &09:55:53.1             &  +69:40:41            &1373.69          &187                 &3.63            &2017 Jul         &106  &145\\
M82\_4               &09:55:55.0             &  +69:40:50             &1373.69          &187                 &3.63            &2019 Feb       &118  &134\\
M82\_5               &09:55:57.0            &  +69:40:55             &1373.69          &187                 &3.63            &2019 Feb       &89  &101\\
NGC 3079         & 10:01:57.9        &+55:40:51            &104.69            &1166                & 16.1           &2017 Jul        &122   &90\\
NGC 3310         &10:38:46.2         &  +53:30:08       &   44.19       & 1060      & 19.81     &     2019  Feb &120   & 83  \\
NGC 3351         &10:43:58.1         &  +11:42:10           &   41.10          &  769            &  9.99     &  2019 Jul      &65  &112 \\
NGC 3368         &10:46:45.6         &  +11:49:13           &   31.63          &  916             & 10.51     &  2019 Jul      &47  &93\\
NGC 3504         &11:03:11.1          &  +27:58:22          &   34.05           & 1550          & 27.07    & 2017 Jul      &67  &99 \\
NGC 3521         &11:05:49.2         &  -00:02:15.0     &  121.76    &  736     &  6.84      &      2019  Feb  & 102    &  90   \\
NGC 3593         &11:14:37.4         &  +12:49:02           &   36.44          &  578              &  5.04     &  2019 Jul      &64  &91\\
NGC 3627         &11:20:15.3    &  +12:59:32    &  136.56    &  740      & 10.04      &     2019  Feb  & 85    &  87   \\
NGC 3628         &11:20:17.4     &  +13:35:19    &  105.76    &  825      &  10.04    &    2019  Feb  & 52     & 90   \\
NGC 3675         &11:26:09.0         &  +43:35:04          &   36.56           &  804         & 12.69      & 2019 Jul      &74   &101 \\
NGC 3690         &11:28:31.0    &  +58:33:48      &  111.42      & 3159      & 47.74    &     2019  Feb   &102   & 82  \\
NGC 3810         &11:40:57.8          &  +11:28:18         &   35.07      & 1001          & 15.36     & 2019 Jul      &63    &106 \\
NGC 3893         &11:48:37.7         &  +48:42:45          &   36.80           &  892         & 16.35       &   2019 Jul      &78   &105 \\
IC 694               &11:28:34.2    &  +58:33:48       &  111.42      & 3100      & 47.74    &     2017 Jul    &120   & 83  \\
NGC 3953         &11:53:48.4     &  +52:19:44       &   31.12      & 1016      & 17.58     & 2019  May     &120   & 115  \\
NGC 4030         &12:00:23.8          &  -01:06:03        &   50.92      & 1427        & 24.50        & 2019 Jul      &81    & 108  \\
NGC 4041         &12:02:12.8      &  +62:08:07     &   31.74        & 1243       & 22.78    & 2019  May      &100   & 92  \\
NGC 4088         &12:05:35.1    &  +50:32:24     &   61.68     &  696        & 13.37    &     2019  Feb  &81   &  98   \\
NGC 4102         &12:06:23.6         &  +52:42:36          &   70.29          &  859        & 16.89    & 2019 Jul      &47   &99 \\
NGC 4157         &12:11:04.2         &  +50:29:04           &   50.67           &  790        & 13.30   & 2019 Jul      &71   &99 \\
NGC 4217         &12:15:50.3       &  +47:05:30      &   41.19          & 1028        & 17.13     &    2019  May      &102   & 90  \\
NGC 4254         &12:18:51.0  &  +14:24:50     &   91.86     & 2403     & 15.29     &    2019  Feb  &  94   & 86    \\
NGC 4303         &12:21:55.4    &  +04:28:24     &   78.74     & 1570      & 15.29    &    2019  Feb   &  83    &  92   \\
NGC 4321         &12:22:53.9        &  +15:49:22        &   68.37       & 1571         & 15.20     &   2019 Jul      &48   & 130  \\
NGC 4414         &12:26:26.9  &  +31:13:24      &   70.69    &  720      & 17.68     &      2019  Feb  & 77      &   85   \\
NGC 4418         &12:26:54.7  &  -00:52:42      &   31.94     & 2104     & 31.90     &   2019  Feb  &  81   &  108    \\
NGC 4501         &12:31:57.6       &  +14:25:20          &   62.97     & 2284         & 15.29      &   2019 Jul      &54  & 109  \\
NGC 4490         &12:30:34.9     &  +41:38:47    &   88.61     &  641        & 10.48   &     2019  Feb  &102    &  90   \\
NGC 4535         &12:34:19.9        &  +08:11:52       &   32.52          & 1957       & 15.77     &  2019 Jul      &67   & 105  \\
NGC 4527         &12:34:09.9        &  +02:39:04     &   65.68     & 1771     & 15.29     &     2019  Feb  &  102   &85   \\
NGC 4536         &12:34:28.5       &  +02:11:08      &   44.51    & 1802     & 14.92     &     2019  Feb  &  138   &85   \\
NGC 4568        &12:36:33.7          &  +11:14:32           &   56.80          & 2262            & 15.29      & 2019 Aug     &37 &112\\   
NGC 4605         &12:40:00.9         &  +61:36:28         &   30.55          &  117         &  3.90     &  2019 Jul      &64    &107 \\
NGC 4631         &12:42:07.1          &  +32:32:33          &  160.08         &  630                 &  7.73       & 2019 Aug     &34  &94\\     
NGC 4654         &12:43:56.6         &  +13:07:39           &   37.77          &  1037           & 15.29     & 2019 Aug    &47  &94\\
NGC 4736         &12:50:52.9          &  +41:07:15          &  120.69          &  323                  &  4.83     &  2019 Aug     &74  & 86\\ 
NGC 4666         &12:45:07.7         &  -00:27:41            &   85.95          & 1495             &  12.82    &  2019 Jul      &41  &99\\
Mrk 231             &12:56:14.2         &+56:52:25             &     29.74        & 12139           & 175         &   2017 Jul      &153  &94\\
NGC 4826         &12:56:42.6          &  +21:41:05          &   81.65           &  349                  &  3.09      &  2019 Aug     &29  & 82\\ 
NGC 5033         &13:13:27.2          &  +36:35:40          &   50.23            &  869                 & 13.76     &  2019 Aug     &34 & 91\\
NGC 5055         &13:15:49.5          &  +42:01:39          &  139.82           &  500                 &  7.96     &  2019 Aug     &40  & 91\\ 
NGC 5054         &13:16:59.0         &  -16:38:04            &   31.53          & 1737            & 23.87      &2019 Jul      &78  &114\\
NGC 5194         &13:29:53.5          &  +47:11:42          &  221.21          &  468                  &  8.63      &  2019 Aug     &81  & 87\\ 
NGC 5195         &13:30:00.0          &  +47:16:00          &   31.33            &  450                 &  8.30     &  2019 Aug     &54  & 90\\ 
NGC 5248         &13:37:31.8         &  +08:53:12         &   53.48       & 1152          & 13.82      & 2019 Jul      &84    &140 \\
NGC 5247         &13:38:03.4         &  -17:53:04            &   41.83          & 1362            & 18.77      &2019 Jul      &58  &112\\
NGC 5713        &14:40:10.9          &  -00:17:22            &   37.28         & 1904            & 26.74       & 2019 Aug    &55  &119\\   
NGC 5457         &14:03:09.0          &  +54:21:24          &  252.84          &  241                 &  6.70        &  2019 Aug     &54  &94\\ 
NGC 5775         &14:53:58.0         &  +03:32:32           &   55.64          & 1642            & 26.34      & 2019 Jul      &47  &94\\
CGCG 049-057 &15:13:12.7         &  +07:13:30           &   31.53          & 3893           & 59.06    &  2019 Jul      &47  &96\\
NGC 5907         &15:15:58.9          &  +56:18:36          &   37.43          &  612               & 12.08      &2019 Aug     &102  &186\\  
Arp 220            & 15:34:57.1         &+23:30:10             &119.25           & 5450            &  79.90     & 2019 Aug      &47  &94\\
NGC 6240        & 16:52:58.6         &+02:24:03             & 26.49           & 7298            &  103.86    & 2017 Jul      &38  &164\\    
NGC 6822         &19:44:56.8         &  -14:48:24           &   95.42          &  -56           &  0.54       &   2019 Jul      &70   &115 \\
NGC 6946         &20:34:52.6    &  +60:09:12      &  290.69    &   53         &  5.32     &     2017   Jul  &51   &  90   \\

\hline
\end{longtable}
\end{center}

\clearpage
\twocolumn

%
%
%


\clearpage
\onecolumn 
 \centering
 \small
\begin{longtable}{llllllllllllllll}

    \caption{Velocity-integrated Intensities}	
     \label{tab:intensity}\\
     		
  	         \hline
  	        \hline
	         
	         {Source Name} & {I$\rm_ {HCN (1-0)}$} & {I$\rm_{HCO^{+}(1-0)}$} & {I$\rm_ {HNC (1-0)}$} & {I$\rm_ {CS (3-2)}$}    \\
	                        & (K km s$^{-1}$)& (K km s$^{-1}$)&(K km s$^{-1}$)&(K km s$^{-1}$)\\
	         \hline
	         
\endfirsthead

\caption[]{(continued)}\\

  	         \hline
  	        \hline
	         
	         {Source Name} & {I$\rm_ {HCN (1-0)}$} & {I$\rm_{HCO^{+}(1-0)}$} & {I$\rm_ {HNC (1-0)}$} & {I$\rm_ {CS (3-2)}$}  \\
	                   & (K km s$^{-1}$)& (K km s$^{-1}$)&(K km s$^{-1}$)&(K km s$^{-1}$)\\
	         \hline

    \endhead
\hline
\endfoot


NGC 891   &   0.63  $\pm$    0.07          &  0.65     $\pm$ 0.08        &   0.73       $\pm$  0.13          &          $<0.29$                          \\
IC 342    &  9.40   $\pm$   0.13           &   7.84    $\pm$  0.11       &     4.75     $\pm$  0.11        &            2.45  $\pm$ 0.13                      \\
UGC 2855  &  2.82   $\pm$   0.11           &  2.44     $\pm$   0.12      &  1.23        $\pm$   0.12       &            0.59   $\pm$  0.14                   \\
NGC 1569   & $<0.14$        &   0.34    $\pm$  0.10       &       $<0.19$         &            $<0.33$                          \\
NGC 2146  & 4.15    $\pm$    0.16          &   5.09    $\pm$   0.27      &     1.19     $\pm$   0.21       &    1.22        $\pm$    0.23               \\
NGC 2403  &   $<0.15$             &   $<0.15$       &     $<0.15$       &       $<0.26$                   \\
NGC 2683   & 0.29    $\pm$     0.06         &   0.12    $\pm$   0.03      &      $<0.13$          &           $<0.19$                     \\
NGC 2903     &  2.69   $\pm$    0.14          &   2.31    $\pm$  0.14       &     0.63     $\pm$   0.14       &            $<0.65$                 \\
NGC 2976   &  0.20   $\pm$   0.07           &    0.10   $\pm$    0.03     &       $<0.14$          &       0.31     $\pm$    0.09                    \\
NGC 3031 &     $<0.18$               &   0.28    $\pm$   0.05      &        $<0.12$        &        $<0.37$                \\
M 82\_1     &    16  $\pm$  0.04    &   24   $\pm$   0.07         &   6.9   $\pm$ 0.06      &   4.94 $\pm$   0.08 \\
M 82\_2     &     26  $\pm$  0.08        &   39.6   $\pm$  0.09       &  12.4    $\pm$  0.08  &     11.35  $\pm$ 0.08\\
M 82\_3      &     27.6   $\pm$   0.07            &  41.5  $\pm$  0.11          &  13.4  $\pm$ 0.08   &  12.93  $\pm$  0.45  \\  
M 82\_4      &  25.8     $\pm$   0.57              &    36  $\pm$   0.11                 &  11.4  $\pm$  0.08         &9.50    $\pm$  0.33\\
M 82\_5      &   15  $\pm$  0.06                &   20   $\pm$ 0.10              &5.7   $\pm$  0.14    &  4.31   $\pm$ 0.13\\
NGC 3079   &   5.6   $\pm$  0.14               &    5.9   $\pm$  0.56         &   2.2  $\pm$  0.14      &  3.12     $\pm$   0.20 \\    
NGC 3310     &  0.38    $\pm$    0.06          &   0.54    $\pm$  0.09   &     $<0.13$   &     0.53       $\pm$   0.06                \\
NGC 3351     &  1.86   $\pm$    0.23          &    1.47    $\pm$    0.15        &       0.84    $\pm$     0.11     &      0.88      $\pm$     0.25            \\
NGC 3368   &    1.82   $\pm$    0.12          & 2.29      $\pm$  0.20       &  0.87        $\pm$    0.16      &    1.43        $\pm$    0.21         \\
NGC 3504    &   2.11  $\pm$      0.10        &   2.04    $\pm$   0.12      &   0.82       $\pm$   0.14       &     0.94       $\pm$   0.15          \\
NGC 3521  & 0.70    $\pm$       0.10            &  0.30     $\pm$  0.10    &       $<0.16$         &          $<0.28$               \\
NGC 3593   &  1.01     $\pm$     0.16         &   1.14    $\pm$   0.11      &    0.75      $\pm$    0.13      &    0.58        $\pm$     0.16        \\
NGC 3627      &    3.26   $\pm$   0.09           &   2.38    $\pm$  0.07      &    1.06      $\pm$   0.05       &    1.34        $\pm$   0.09          \\
NGC 3628      &   4.32  $\pm$    0.16          &  4.30     $\pm$  0.17   &  2.83        $\pm$   0.07       &   2.09         $\pm$     0.17          \\
NGC 3675    &  0.68   $\pm$    0.17          &   0.37    $\pm$    0.12     &       $<0.29$           &         $<0.40$            \\
NGC 3690     &1.07     $\pm$    0.08          &  2.17     $\pm$  0.11   &   0.80       $\pm$     0.07     &    0.68        $\pm$    0.09          \\
NGC 3810   &  0.31   $\pm$    0.10          &   $<0.18$         &    0.47      $\pm$    0.11      &   0.69         $\pm$    0.24            \\
NGC 3893     &   0.76  $\pm$     0.13         &   0.54    $\pm$   0.11      &         $<0.22$       &          $<0.36$                  \\
IC 694            &  2.66   $\pm$     0.10         &   4.69    $\pm$ 0.11    &    0.74      $\pm$  0.11        &    2.01        $\pm$    0.22           \\
NGC 3953     & 0.31    $\pm$    0.07          &   0.25    $\pm$    0.05   &       $<0.11$      &          $<0.15$             \\
NGC 4030   &  2.23   $\pm$     0.11         &    1.36   $\pm$    0.12     &  0.91        $\pm$      0.14    &    0.51        $\pm$    0.13          \\
NGC 4041     & 2.15    $\pm$   0.07           &   1.52    $\pm$   0.04    &   0.85      $\pm$    0.06      &  0.50          $\pm$     0.07              \\
NGC 4088     &  0.92   $\pm$    0.05          &1.09     $\pm$0.07     &   0.25       $\pm$   0.07       &          $<0.33$               \\
NGC 4102     & 5.16    $\pm$    0.14          &   3.94    $\pm$   0.16      &   2.51       $\pm$   0.23       &    1.42        $\pm$   0.25           \\
NGC 4157    &  0.54   $\pm$     0.13         &   0.89    $\pm$  0.27       &      $<0.19$    &           $<0.33$                 \\
NGC 4217   &   0.61  $\pm$      0.07        &   0.40    $\pm$  0.09   &    0.33        $\pm$  0.11      &          $<0.18$            \\
NGC 4254     &   0.60 $\pm$ 0.04    & 0.50$\pm$   0.06    & 0.23$\pm$ 0.07      & $<0.28$  \\
NGC 4303       &  1.41   $\pm$   0.09          &  1.20     $\pm$  0.06       &  0.55     $\pm$   0.07       &   0.76         $\pm$     0.08              \\
NGC 4321   &  1.99   $\pm$      0.12        &   1.28    $\pm$     0.18   &    1.08      $\pm$   0.20       &        $<0.53$             \\
NGC 4414     &   2.57    $\pm$    0.21                      & 1.57 $\pm$ 0.14     &  1.04   $\pm$ 0.04          &   $<0.23$     \\
NGC 4418     &   2.48 $\pm$ 0.13      & 1.89$\pm$   0.11      & 1.37$\pm$  0.10       &  1.91  $\pm$  0.15  \\
NGC 4501   &  0.92    $\pm$  0.11            & 0.58      $\pm$   0.15      &   0.32       $\pm$    0.09      &   0.48         $\pm$    0.15        \\
NGC 4490      &  0.28     $\pm$    0.09       &   0.19  $\pm$ 0.04     &     0.19  $\pm$ 0.04     &            $<0.23$             \\
NGC  4535   & 0.90    $\pm$    0.06          &    0.41   $\pm$  0.07       &   0.35       $\pm$   0.08         &   0.60     $\pm$     0.10        \\
NGC 4527     &      $<0.10$                          & 0.21$\pm$  0.07     & 0.21$\pm$  0.07  & $<0.22$   \\
NGC 4536     &           $<0.1$              & 0.23$\pm$  0.05     &$<0.07$  &  $<0.07$  \\
NGC 4568   &0.99 $\pm$       0.14       &  0.94     $\pm$    0.20     &        $<0.37$      &            $<0.59$                   \\
NGC 4605         & $<0.19$            &     $<0.22$        &      $<0.10$        &           $<0.24$                 \\
NGC 4631   &  0.64   $\pm$    0.06          &  1.20     $\pm$   0.08      &      $<0.24$     &         $<0.31$             \\
NGC 4654    &    $<0.22$              &     $<0.26$        &       $<0.27$     &          $<0.26$             \\
NGC 4736 &   0.91  $\pm$     0.09         &    0.88   $\pm$   0.11      &   0.50       $\pm$     0.08     &      0.52      $\pm$     0.12          \\
NGC 4666  &  1.41   $\pm$  0.19            & 1.04      $\pm$   0.14      &   0.48       $\pm$   0.11       &            $<0.48$               \\
Mrk 231      &  1.6   $\pm$ 0.05                &   1.2   $\pm$  0.05         &  0.7  $\pm$   0.03          &   $-$     \\
NGC 4826 &   4.08  $\pm$     0.12         &   2.91    $\pm$    0.17     &     1.57     $\pm$   0.19       &   0.64         $\pm$     0.16         \\
NGC 5033 &   1.76  $\pm$       0.16       & 1.76      $\pm$      0.19   &   0.75       $\pm$  0.31        &    0.37        $\pm$   0.12          \\
NGC 5055   &  1.71   $\pm$    0.19          &   1.37    $\pm$  0.29       &     0.38     $\pm$   0.13       &       $<0.34$             \\
NGC 5054   &   0.64  $\pm$   0.12           &   0.74    $\pm$    0.18     &  0.29        $\pm$    0.09      &   0.75         $\pm$    0.14           \\
NGC 5194 &  4.59   $\pm$       0.14       &   2.59    $\pm$  0.12       &    1.27      $\pm$   0.07       &        $<0.23$               \\
NGC 5195  &   1.40  $\pm$      0.14        &   2.36    $\pm$   0.14      &    0.78      $\pm$     0.11     &          $<0.24$             \\
NGC 5248   &  2.22   $\pm$   0.18           &  1.17     $\pm$  0.16       &   0.99       $\pm$     0.15     &          $<0.30$                \\
NGC 5247     &   0.42  $\pm$    0.09          & 1.24      $\pm$  0.25       &       $<0.18$         &        $<0.31$                \\
NGC 5713   & 1.17    $\pm$   0.11           &   1.05    $\pm$  0.10       &   0.45       $\pm$ 0.09         &   0.45         $\pm$      0.13         \\
NGC 5457 & 0.22    $\pm$    0.04          &    $<0.12$       &     $<0.17$          &           $<0.22$              \\
NGC 5775   & 0.60 $\pm$ 0.12           &   0.59$\pm$ 0.08    &  0.27  $\pm$ 0.13    &$<0.58$\\
CGCG049-057  &   2.35  $\pm$    0.21          &  1.35     $\pm$    0.19     &    0.88      $\pm$     0.11     &     1.14       $\pm$    0.20            \\
NGC 5907   &  0.15   $\pm$ 0.05        &     $<0.12$        &    0.24      $\pm$   0.04       &          $<0.30$           \\
Arp 220    &  8.58   $\pm$   0.31         &   3.64    $\pm$  0.30      &    6.73      $\pm$    0.31      &     4.97       $\pm$    0.31            \\
NGC 6240   &    2.7 $\pm$    0.05          &   2.77    $\pm$    0.40     &        0.60  $\pm$0.04       &         $-$           \\
NGC 6822    &$  <0.31  $        &      $<0.25$        &       $<0.23$     &          $<0.26$            \\
NGC 6946     &  9.34   $\pm$      0.10        &    9.05   $\pm$ 0.13    &   3.60       $\pm$   0.30       &   3.61         $\pm$    0.11       \\


\hline	

 		

\end{longtable}

\clearpage
\twocolumn

\clearpage
\onecolumn 
 \small
\begin{longtable}{llllllllllllllll}

    \caption{Line luminosities and infrared luminosity}	
     \label{tab:luminosity}\\
     		
  	         \hline
  	        \hline
	         
	         {Source} & {log$L\rm_{HCN}$} & { log$L\rm_{HCO^+}$} &  { log$L\rm_{HNC}$} &{ log$L\rm_{ CS}$}& { {log$L\rm_{IR}$}$^{\rm a}$  }  &  { {log$L\rm_{IR}$}$^{\rm b}$  } & {Type}   \\
	                   & (K km s$^{-1}$)& (K km s$^{-1}$)&(K km s$^{-1}$)&(K km s$^{-1}$)&log$L_{\odot}$& log$L_{\odot}$ &\\

	         \hline
	         
\endfirsthead
\caption[]{(continued)}\\

  	         \hline
  	        \hline
	       {Source} & {log$L\rm_{ HCN}$} & { log$L\rm_{HCO^+}$} &  { log$L\rm_{ HNC}$} &{ log$L\rm_{ CS}$}& { {log$L\rm_{IR}$} }  & { {log$L\rm_{IR}$} }&  {Type}   \\
	                   & (K km s$^{-1}$)& (K km s$^{-1}$)&(K km s$^{-1}$)&(K km s$^{-1}$)&(log$L_{\odot}$)& (log$L_{\odot}$) &\\

	         \hline

    \endhead
\hline
\endfoot

NGC 891 	&	5.98 	$\pm$	0.05 	&	5.99 	$\pm$	0.05 	&	6.04 	$\pm$	0.08 &$<5.64$	&	9.89	$\pm$0.11	& 9.69$\pm$0.11	&  SA(s)b	\\
IC 342  	&	6.61 	$\pm$	0.01 	&	6.53 	$\pm$	0.01 	&	6.32 	$\pm$	0.01  &  6.02$\pm$ 0.02	&	10.14	$\pm$0.09	&10.07$\pm$0.09  &	SAB(rs)cd	\\
$^*$UGC 2855	&	7.34 	$\pm$	0.02 	&	7.28 	$\pm$	0.02 	&	6.98 	$\pm$	0.04  &0.66$\pm$ 0.10	&	10.75$\pm$    0.09  & 10.75$\pm$0.09	&	SB	\\
NGC 1569	&	$<4.79 	$		&	5.17 	$\pm$	0.13 	&$<	5.23 $	&$<5.16$  &	8.86	$\pm$  0.11     	&8.36$\pm$0.11  &	IBm	\\
NGC 2146	&	7.36 	$\pm$	0.02 	&	7.45 	$\pm$	0.02 	&	6.82 	$\pm$	0.08 &	 6.83 $\pm$ 0.08 &	10.50	$\pm$0.11	& 9.55$\pm$0.11 &	SB	\\
NGC 2403	&$<	4.51 	$		&$<	4.51 $	 	&	$<4.28 	$	&$<4.74$ 	&	7.64	$\pm$ 0.11	&	7.04$\pm$0.11 &SAB(s)cd	\\
NGC 2683	&	5.51 	$\pm$	0.09 	&	5.12 	$\pm$	0.11 	&$<	5.16 	$	&$<5.32$	&	8.68	$\pm$0.11	& 8.24$\pm$0.09  &	SA(rs)b	\\
NGC 2903 	&	6.58 	$\pm$	0.02 	&	6.51 	$\pm$	0.03 	&	5.95 	$\pm$	0.10 & $<5.96$	&	9.83	$\pm$0.09	& 9.73$\pm$0.09   &	SAB(rs)bc	\\
NGC 2976	&	4.47 	$\pm$	0.15 	&	4.17 	$\pm$	0.13 	&$<	4.32 	$	&4.66$\pm$ 	0.13	&	7.59	$\pm$0.09	&7.23$\pm$0.11&	SAc	\\
NGC 3031	&$<	4.69 $			&	4.88 	$\pm$	0.08 	&	$<4.51 	$	&$<5.00$	&	8.23	$\pm$0.11	& 8.03$\pm$0.11 &	SA(s)ab	\\
M 82\_1 	&	6.64 $\pm$0.001	&	6.81 $\pm$	0.001	&	6.27$\pm$	0.004 	&6.13	$\pm$	0.01&	10.04 $\pm$0.11	&9.65$\pm$0.11 &	SB	\\
M 82\_2 	&	6.85 $\pm$	0.001	&	7.03 $\pm$	0.001	&	6.53 $\pm$0.003	&	6.49	$\pm$	0.01&	10.33$\pm$0.11 & 10.11$\pm$0.11	&	SB	\\
M 82\_3 	&	6.87$\pm$ 	0.001		&	7.05 $\pm$	0.001		&	6.56$\pm$0.003 	&	6.55	$\pm$	0.02	&	10.31$\pm$0.11	&9.93$\pm$0.11 &	SB	\\
M 82\_4 	&	6.85 	$\pm$	0.01 	&	6.99 $\pm$	0.001		&	6.49$\pm$0.003 	&	6.41	$\pm$		0.02	&	10.27$\pm$0.11	& 10.10$\pm$0.11&	SB	\\
M 82\_5 	&	6.61 $\pm$ 	0.001		&	6.73 $\pm$	0.002		&	6.19 $\pm$0.01	$\pm$	0.01 &	6.07	$\pm$	0.01 &	9.95	$\pm$0.12&	9.50$\pm$0.11 & SB	\\
NGC 3079	&	7.58 	$\pm$	0.01 	&	7.60 	$\pm$	0.04 	&	7.17 	$\pm$	0.03   &  7.32$\pm$	  0.03&	10.77$\pm$0.11	& 10.70$\pm$0.11 &	SB	\\
NGC 3310	&	6.48 	$\pm$	0.07 	&	6.64 	$\pm$	0.07 	&$<		6.02 	$	  &6.63$\pm$	  0.05	&	10.54$\pm$0.09	& 10.44$\pm$0.11&	SB	\\
NGC 3351	&	6.58 	$\pm$	0.05 	&	6.48 	$\pm$	0.04 	&	6.23 	$\pm$	0.11  & 6.25$\pm$	  0.12  &	9.75$\pm$0.11	& 	9.58$\pm$0.11 & SB(r)b	\\
NGC 3368	&	6.61 	$\pm$	0.03 	&	6.71 	$\pm$	0.04 	&	6.29 	$\pm$	0.08  & 6.51$\pm$	  0.06  	&	9.26$\pm$0.09	& 9.15$\pm$0.09 & 	SAB(rs)ab	\\
NGC 3504	&	7.50 	$\pm$	0.02 	&	7.48 	$\pm$	0.03 	&	7.09 	$\pm$	0.07 &7.15	$\pm$0.07 &	10.53$\pm$0.11	&10.44$\pm$0.11&	SB	\\
NGC 3521	&	5.83 	$\pm$	0.06 	&	5.46 	$\pm$	0.14 	&	$<	5.19 	$	&$<5.43$	&	9.18$\pm$0.11	&	8.70$\pm$0.11& SAB(rs)bc	\\
NGC 3593	&	5.72 	$\pm$	0.26 	&	5.77 	$\pm$	0.04 	&	5.59 	$\pm$	0.08 & 5.48$\pm$0.12	&	9.17$\pm$0.09	& 9.03$\pm$0.09&	SA(s)	\\
NGC 3627	&	6.83 	$\pm$	0.01 	&	6.69 	$\pm$	0.01 	&	6.34 	$\pm$	0.02 &6.44$\pm$0.03	&	9.73$\pm$0.11	& 9.54$\pm$0.11&	SB	\\
NGC 3628	&	6.95 	$\pm$	0.02 	&	6.95 	$\pm$	0.02 	&	6.77 	$\pm$	0.01 &	6.64$\pm$0.04 &	10.28$\pm$0.11	&10.13$\pm$0.11&	SB	\\
NGC 3675	&	6.35 	$\pm$	0.11 	&	6.09 	$\pm$	0.14 	&	$<	5.98 	$	&$<6.12$	&	9.50	$\pm$0.09& 9.26$\pm$0.09&	SA(s)b	\\
NGC 3690	&	7.69 	$\pm$	0.03 	&	8.00 	$\pm$	0.02 	&	7.56 	$\pm$	0.04 &	7.49 $\pm$0.06 &	11.57$\pm$0.12	&	10.13 $\pm$0.11& SB	\\
$^*$NGC 3810	&	6.17 	$\pm$	0.14 	&$<		5.94 $			&	6.36 	$\pm$	0.10 &6.52$\pm$0.15	&	10.12$\pm$0.09	&	10.12$\pm$0.11 &SB	\\
$^*$NGC 3893	&	6.62 	$\pm$	0.07 	&	6.47 	$\pm$	0.09 	&$<		6.08 $		&$<6.30$	&	10.20$\pm$0.09	& 10.20$\pm$0.09	&SB	\\
IC 694  	&	8.08 	$\pm$	0.02 	&	8.33 	$\pm$	0.01 	&	7.52 	$\pm$	0.06 &7.96$\pm$0.05	&	11.74$\pm$0.11	&11.65$\pm$0.11&	SB	\\
NGC 3953 &6.29$\pm$0.10    &  6.20$\pm$0.09   &  $<5.84$   &$<5.98$   & 8.96 $\pm$0.09  &8.77$\pm$0.11 &SB(r)bc    \\
NGC 4030	&	7.43 	$\pm$	0.02 	&	7.22 	$\pm$	0.04 	&	7.05 	$\pm$	0.07 	& 6.79$\pm$0.11  &	10.35$\pm$0.11	& 10.03$\pm$0.09&	SB	\\
NGC 4041	&	7.36 	$\pm$	0.01 	&	7.21 	$\pm$	0.01 	&	6.95 	$\pm$	0.03 &	6.72$\pm$0.06 &	10.35$\pm$0.11	& 10.19$\pm$0.09 &	SB	\\
NGC 4088	&	6.53 	$\pm$	0.02 	&	6.60 	$\pm$	0.03 	&	5.96 	$\pm$	0.12 &$<6.08$	&	9.76$\pm$0.11	& 9.48$\pm$0.09&	SB	\\
NGC 4102	&	7.48 	$\pm$	0.01 	&	7.36 	$\pm$	0.02 	&	7.17 	$\pm$	0.04 &6.92$\pm$0.08	&	10.46$\pm$0.11	& 10.35$\pm$0.11&	SB/AGN	\\
NGC 4157	&	6.29 	$\pm$	0.10 	&	6.51 	$\pm$	0.13 	&	5.84 	$\pm$	0.00 &$<6.08	$&	9.66$\pm$0.09 &	9.42 $\pm$0.09&SB	\\
NGC 4217	&	6.56 	$\pm$	0.05 	&	6.38 	$\pm$	0.10 	&	6.30 	$\pm$	0.14 &$<6.03	$	&	9.83$\pm$0.11	&	9.56$\pm$0.11& SB	\\
NGC 4254	&	6.45 	$\pm$	0.03 	&	6.37 	$\pm$	0.05 	&	6.03 	$\pm$	0.13 &	$<6.12	$&	9.78$\pm$0.11	&	9.56 $\pm$0.11&SB	\\
NGC 4303	&	6.83 	$\pm$	0.03 	&	6.76 	$\pm$	0.02 	&	6.42 	$\pm$	0.06 &6.56$\pm$0.05	&	9.60$\pm$0.11	&	9.46$\pm$0.11 &SB	\\
NGC 4321	&	6.97 	$\pm$	0.03 	&	6.78 	$\pm$	0.06 	&	6.71 	$\pm$	0.08 &$<6.40	$	&	9.73$\pm$0.11	& 9.53$\pm$0.09 &	SB	\\
NGC 4414	&	7.22 	$\pm$	0.04 	&	7.00 	$\pm$	0.04 	&	6.82 	$\pm$	0.02 &$<6.17$	&	10.26$\pm$0.09	& 9.88$\pm$0.09 &	SB	\\
NGC 4418	&	7.71 	$\pm$	0.02 	&	7.59 	$\pm$	0.03 	&	7.45 	$\pm$	0.03 &	7.60$\pm$0.03&	10.99$\pm$0.09	&	10.98$\pm$0.11 &SB/AGN	\\
NGC 4501	&	6.64 	$\pm$	0.05 	&	6.44 	$\pm$	0.11 	&	6.18 	$\pm$	0.12 &6.35$\pm$0.14	&	9.56$\pm$0.09	&	9.23$\pm$0.11 &SB/AGN	\\
NGC 4490	&	5.75 	$\pm$	0.14 	&	5.58 	$\pm$	0.09 	&	5.58 	$\pm$	0.09 &	$<5.71$&	9.70$\pm$0.09	& 9.41$\pm$0.09&	SB	\\
NGC 4535	&	6.66 	$\pm$	0.03 	&	6.31 	$\pm$	0.07 	&	6.25 	$\pm$	0.10 &	6.48$\pm$0.07&	9.43$\pm$0.11	&	9.23$\pm$0.09& SB	\\
NGC 4527	&$<		5.68 	$		&	6.00 	$\pm$	0.14 	&	6.00 	$\pm$	0.14 &$<6.02	$&	9.55$\pm$0.11	&	8.68$\pm$0.11 &SB	\\
NGC 4536	&$<		5.65 	$		&	6.02 	$\pm$	0.09 	&	$<	5.56 	$	&$<5.50$	&	9.67$\pm$0.11	& 9.86$\pm$0.11&	SB	\\
NGC 4568	&	6.67 	$\pm$	0.06 	&	6.65 	$\pm$	0.09 	&$<		6.25 	$	&$<6.44$	&	9.28$\pm$0.11	& 8.77$\pm$0.11&	SB	\\
NGC 4605	&	4.78 	$\pm$	0.00 	&	4.84 	$\pm$	0.00 	&$<		4.50 	$	&$<4.88$	&	8.35$\pm$0.09& 7.95$\pm$0.09&	SB(s)c	\\
NGC 4631	&	5.89 	$\pm$	0.04 	&	6.17 	$\pm$	0.03 	&$<		5.47 	$	&$<5.58$	&	9.54$\pm$0.11	&	9.23$\pm$0.11 &SB	\\
NGC 4654	&$<		6.02 	$		&$<		6.09 $		&	$<	6.11 	$  &	$<6.09$	&	9.59$\pm$0.11	&	9.30$\pm$0.11 &SB	\\
NGC 4736	&	5.64 	$\pm$	0.04 	&	5.63 	$\pm$	0.05 	&	5.38 	$\pm$	0.07 & 5.40	$\pm$0.10&	9.20$\pm$0.11	&	8.96$\pm$0.11 & AGN	\\
NGC 4666	&	6.67 	$\pm$	0.06 	&	6.54 	$\pm$	0.06 	&	6.21 	$\pm$	0.10 &$<6.21$	&	9.74$\pm$0.11	& 9.15$\pm$0.11 &	SB	\\
Mrk 231 	&	8.94 	$\pm$	0.01 	&	8.81 	$\pm$	0.03 	&	8.58 	$\pm$	0.02 & $-$	&	12.39$\pm$0.11	&	8.96$\pm$0.11&SB	\\
NGC 4826	&	5.90 	$\pm$	0.01 	&	5.76 	$\pm$	0.03 	&	5.49 	$\pm$	0.05 &5.10	$\pm$0.12&	8.80$\pm$0.11	& 8.42$\pm$0.11 &	AGN	\\
NGC 5033	&	6.83 	$\pm$	0.04 	&	6.83 	$\pm$	0.05 	&	6.46 	$\pm$	0.18 &	6.17$\pm$0.14&	9.29$\pm$0.11	&	9.01$\pm$0.11&SB/AGN	\\
NGC 5055	&	6.35 	$\pm$	0.05 	&	6.25 	$\pm$	0.09 	&	5.69 	$\pm$	0.15 &$<5.65$	&	9.42$\pm$0.11	& 9.01$\pm$0.11&	SB	\\
$^*$NGC 5054	&	6.87 	$\pm$	0.08 	&	6.93 	$\pm$	0.11 	&	6.53 	$\pm$	0.13 &6.94$\pm$0.08	&	10.34$\pm$0.09	&10.34$\pm$0.09 &SB	\\
NGC 5194	&	6.85 	$\pm$	0.01 	&	6.60 	$\pm$	0.02 	&	6.29 	$\pm$	0.02 &	$<5.55$&	9.53$\pm$0.11&	9.07$\pm$0.11&SB/AGN	\\
NGC 5195	&	6.30 	$\pm$	0.04 	&	6.52 	$\pm$	0.03 	&	6.04 	$\pm$	0.06 &$<5.53$	&	9.30$\pm$0.12	&	9.21$\pm$0.11&I0	\\
NGC 5248	&	6.94 	$\pm$	0.04 	&	6.66 	$\pm$	0.06 	&	6.59 	$\pm$	0.07 &	$<6.07$&	9.79$\pm$0.11	&	9.54 $\pm$0.11&SB	\\
NGC 5247	&	6.48 	$\pm$	0.09 	&	6.95 	$\pm$	0.09 	&	6.11 	$\pm$	0.00 &$<6.35$	&	9.68$\pm$0.09	& 9.50$\pm$0.09 &	SB	\\
NGC 5713	&	7.23 	$\pm$	0.04 	&	7.18 	$\pm$	0.08 	&	6.81 	$\pm$	0.87 &	6.81$\pm$0.13&	10.54$\pm$0.11	&	10.33$\pm$0.11 &SB	\\
NGC 5457	&	5.31 	$\pm$	0.08 	&	$<	5.04 	$		&$<		5.20 $		&	$<5.31$&	7.96$\pm$0.14	& 7.40 $\pm$0.11&	SB	\\
NGC 5775	&	6.93 	$\pm$	0.09 	&	6.92 	$\pm$	0.07 	&$<		6.60 	$	&	$<6.91$&	10.32$\pm$0.09	&	10.10$\pm$0.09&SB	\\
CGCG049-057	&	8.21 	$\pm$	0.04 	&	7.97 	$\pm$	0.06 	&	7.78 	$\pm$	0.05 	& 7.90$\pm$0.08&	11.34$\pm$0.11	&	11.20$\pm$0.11 &SB	\\
NGC 5907	&	5.65 	$\pm$	0.14 	&$<		5.56 	$		&	5.86 	$\pm$	0.07 &$<5.95	$&	9.06$\pm$0.11& 8.63$\pm$0.11 &	SA(s)c	\\
Arp 220 	&	9.03 	$\pm$	0.02 	&	8.66 	$\pm$	0.04 	&	8.92 	$\pm$	0.02 &8.67	$\pm$0.04 &	12.19$\pm$0.11	&	12.16$\pm$0.11 &SB/AGN	\\
NGC 6240	&	8.75 	$\pm$	0.01 	&	8.76 	$\pm$	0.06 	&	8.09 	$\pm$	0.03 & $-$	&	11.83$\pm$0.11	&	11.78$\pm$0.11 & SB/AGN	\\
NGC 6822	&	$<3.27$	&	$<3.18$ 	&$<		3.14 $		&$<3.19	$	&	5.28$\pm$0.15	&4.78$\pm$0.11 &	IB(s)m	\\
NGC 6946	&	6.74 	$\pm$	0.00 	&	6.72 	$\pm$	0.01 	&	6.32 	$\pm$	0.04 &	5.98$\pm$0.03&	9.63$\pm$0.11	&	9.53$\pm$0.11 &SB	\\
\hline
IRAS17208	&	9.11 	$\pm$	0.03 	&	9.00 	$\pm$	0.04 	&	9.00 	$\pm$	0.05 	&$-$&	12.46$\pm$0.11	&$-$&	SB	\\
IC 860	&	7.96 	$\pm$	0.05 	&	7.76 	$\pm$	0.06 	&	7.69 	$\pm$	0.07 	&$-$&	11.19$\pm$0.11	&	$-$&SB	\\
NGC 1614	&	7.86 	$\pm$	0.16 	&	8.13 	$\pm$	0.09 	&	7.33 	$\pm$	0.42 &$-$	&	11.51$\pm$0.11	&	$-$&SB	\\
NGC 4388	&	6.46 	$\pm$	0.08 	&	6.61 	$\pm$	0.06 	&	6.09 	$\pm$	0.18 	&$-$&	9.52	$\pm$0.11&$-$&	SB/AGN	\\
NGC 6090	&	8.20 	$\pm$	0.07 	&	8.43 	$\pm$	0.04 	&	7.60 	$\pm$	0.22 	&$-$&	11.40$\pm$0.12	&$-$&	SB	\\
NGC 7469	&	8.29 	$\pm$	0.02 	&	8.34 	$\pm$	0.02 	&	8.03 	$\pm$	0.04 	&$-$&	11.51$\pm$0.11	&$-$&	SB/AGN	\\
NGC 7771	&	8.41 	$\pm$	0.02 	&	8.37 	$\pm$	0.02 	&	8.18 	$\pm$	0.04 	&$-$&	11.29$\pm$0.11	&$-$&	SB	\\
NGC 660	&	7.24 	$\pm$	0.02 	&	7.26 	$\pm$	0.02 	&	6.96 	$\pm$	0.03 	&$-$&	9.86$\pm$0.09	&$-$&	SB/AGN	\\
NGC 3556	&	6.39 	$\pm$	0.06 	&	6.59 	$\pm$	0.04 	&	5.87 	$\pm$	0.21 	&$-$&	9.31$\pm$0.09	&$-$&	SB	\\
NGC 1068	&	7.99 	$\pm$	0.01 	&	7.82 	$\pm$	0.01 	&	7.60 	$\pm$	0.02 	&$-$&	10.77$\pm$0.11	&	$-$&SB/AGN	\\
$^*$UGC 2866	&	7.15 	$\pm$	0.04 	&	7.32 	$\pm$	0.02 	&	6.86 	$\pm$	0.07 	&$-$&	10.68$\pm$0.09	&$-$&	SB	\\
NGC 2273	&	7.02 	$\pm$	0.08 	&	6.80 	$\pm$	0.13 	&	6.80 	$\pm$	0.13 	&$-$&	10.18$\pm$0.11&$-$&	SB/AGN	\\
\hline
NGC 34                   	&	7.92 	$\pm$	0.07 	&	8.13 	$\pm$	0.06 	&	$<7.71 	$	&$-$&	11.44$\pm$0.09	&$-$&	SB/AGN	\\
MCG --02-01-052      	&	8.08 	$\pm$	0.10 	&	$<7.76 	$		&	$<7.74 	$	&$-$	&	11.41$\pm$0.09	&$-$&	SB	\\
MCG --02-01-051      	&	7.56 	$\pm$	0.13 	&7.81 	$\pm$	0.10 	&	$<7.53 	$	&$-$	&	11.41$\pm$0.09	&$-$&	SB	\\
IC 1623               	&	8.26 	$\pm$	0.04 	&	8.68 	$\pm$	0.02 	&	7.88 	$\pm$	0.09 &$-$	&	11.65$\pm$0.09	&$-$&	SB	\\
MCG --03-04-014      	&	8.51 	$\pm$	0.06 	&	8.51 	$\pm$	0.04 	&	8.12 	$\pm$	0.14 	&$-$&	11.63$\pm$0.09	&$-$&	SB	\\
IRAS 01364-1042      	&	$<8.65 	$		&$<	8.46 	$	&	$<8.54 	$	&$-$	&	11.76$\pm$0.09	&$-$&	SB/AGN	\\
IC 214               	&	7.97 	$\pm$	0.12 	&	8.16 	$\pm$	0.09 	&	$<7.56 	$	&$-$	&	11.37$\pm$0.09	&$-$&	SB	\\
NGC 0958             	&	$<7.91 $			&	$<7.82 	$		&	$<7.89 	$	&$-$	&	11.17$\pm$0.09	&$-$&	SB	\\
ESO 550-IG 025       	&	8.13 	$\pm$	0.11 	&	8.03 	$\pm$	0.13 	&	$<8.07 	$	&$-$	&	11.45$\pm$0.09	&$-$&	SB	\\
UGC 3094            	&	$<7.88 	$		&	8.13 	$\pm$	0.08 	& $<	7.87 	$	&$-$	&	11.35$\pm$0.09	&	$-$&SB	\\
NGC 1797             	&	7.70 	$\pm$	0.10 	&	7.58 	$\pm$	0.09 	&	$<7.47 	$	&$-$	&	11.00$\pm$0.09	&	$-$&SB	\\
VII Zw 31           	&	$<8.90 	$		&	9.01 	$\pm$	0.14 	&	$<8.89 	$	&$-$	&	11.94$\pm$0.09	&$-$&	SB	\\
IRAS F05189-2524    	&	8.58 	$\pm$	0.08 	&	8.45 	$\pm$	0.10 	&	8.30 	$\pm$	0.12 &$-$	&	12.11$\pm$0.09	&	$-$&SB/AGN	\\
IRAS F05187-1017    	&	8.25 	$\pm$	0.06 	&	8.26 	$\pm$	0.08 	& $<	7.63 		$ &$-$	&	11.23$\pm$0.09	&	$-$&SB	\\
IRAS F06076-2139    	&	8.63 	$\pm$	0.08 	& $<	8.38 	$		& $<	8.31 $	  &$-$ &	11.59$\pm$0.09	&$-$&	SB	\\
NGC 2341            	&	7.57 	$\pm$	0.00 	&	$<7.66 	$		&$<	7.81 	$	&$-$	&	11.25$\pm$0.09	&$-$&	SB	\\
NGC 2342            	&	7.84 	$\pm$	0.07 	&	7.88 	$\pm$	0.08 	&	$<7.35 	$	&$-$	&	11.25$\pm$0.09	&$-$&	SB	\\
IRAS 07251+0248	&	8.82 	$\pm$	0.14 	&$<	8.80 		$	&	$<8.88 	$	&$-$	&	12.32$\pm$0.09	&	$-$&SB	\\
NGC 2623            	&	8.49 	$\pm$	0.08 	&	8.45 	$\pm$	0.10 	&	$<8.16 	$	&$-$	&	11.54$\pm$0.09	&	$-$&SB	\\
IRAS 09111-1007W    	&	8.70 	$\pm$	0.06 	&	8.84 	$\pm$	0.08 	&	$<8.32 	$	&$-$	&	12.00$\pm$0.09	&	$-$&SB/AGN	\\
IRAS 09111-1007E    	&	$<8.55 $			&$<	8.61 $			&	$<8.54 	$	&$-$	&	12.00$\pm$0.09	&	$-$&SB	\\
UGC 5101           	&	9.00 	$\pm$	0.04 	&	8.76 	$\pm$	0.06 	&	8.63 	$\pm$	0.09 &$-$	&	11.95$\pm$0.09	&	$-$&SB/AGN	\\
CGCG 011-076        	&	8.37 	$\pm$	0.06 	&	8.41 	$\pm$	0.06 	&	$<7.88 	$	&$-$	&	11.37$\pm$0.09	&	$-$&SB	\\
IRAS F12224-0624    	&	$<8.13 	$		&$<	8.21 $			&$<	8.21 	$	&$-$	&	11.27$\pm$0.09	&	$-$&SB	\\
CGCG 043-099        	&	$<8.39 	$		&$<	8.38 $			&$<	8.38 	$	&$-$	&	11.62$\pm$0.09	&	$-$&SB	\\
ESO 507-G 070       	&	8.34 	$\pm$	0.07 	&	8.52 	$\pm$	0.05 	&	7.94 	$\pm$	0.12 &$-$	&	11.49$\pm$0.09	&	$-$&SB/AGN	\\
NGC 5104            	&	8.21 	$\pm$	0.08 	&	7.99 	$\pm$	0.12 	&$<	7.74 	$	&$-$	&	11.2	$\pm$0.09&	$-$&SB	\\
IC 4280              	&	8.02 	$\pm$	0.10 	&	8.11 	$\pm$	0.09 	&	7.80 	$\pm$	0.14 &$-$	&	11.08$\pm$0.09	&	$-$&SB	\\
NGC 5257            	&	7.86 	$\pm$	0.11 	&	7.94 	$\pm$	0.08 	&	7.84 	$\pm$	0.13 	&$-$&	11.55$\pm$0.09	&	$-$&SB	\\
NGC 5258             	&	7.91 	$\pm$	0.10 	&	8.00 	$\pm$	0.06 	&	7.75 	$\pm$	0.12 	&$-$&	11.55$\pm$0.09	&	$-$&SB	\\
UGC 8739            	&	8.15 	$\pm$	0.06 	&	8.01 	$\pm$	0.09 	&	7.98 	$\pm$	0.09 &$-$	&	11.08$\pm$0.09	&	$-$&SB	\\
NGC 5331            	&	8.39 	$\pm$	0.06 	&	8.21 	$\pm$	0.10 	&$<	8.19 	$	&$-$	&	11.59$\pm$0.09	&	$-$&SB	\\
CGCG 247-020        	&	8.33 	$\pm$	0.07 	&	8.10 	$\pm$	0.09 	&	7.89 	$\pm$	0.11 &$-$	&	11.32$\pm$0.09	&	$-$&SB	\\
IRAS F14348-1447    	&	8.97 	$\pm$	0.00 	&$<	8.82 $			&$<	8.96 $		&$-$	&	12.30$\pm$0.09	&	$-$&SB	\\
CGCG 049-057         	&	8.35 	$\pm$	0.04 	&	8.01 	$\pm$	0.06 	&	8.07 	$\pm$	0.07 &$-$	&	11.27$\pm$0.09	&	$-$&SB	\\
NGC 5936            	&	7.95 	$\pm$	0.06 	&	7.77 	$\pm$	0.06 	&$<7.53 	$	&$-$	&	11.07$\pm$0.09	&	$-$&SB	\\
ARP 220             	&	9.18 	$\pm$	0.03 	&	8.88 	$\pm$	0.05 	&	8.99 	$\pm$	0.03 &$-$	&	12.21$\pm$0.09	&$-$&	SB/AGN	\\
IRAS F16164-0746    	&	8.14 	$\pm$	0.09 	&	8.38 	$\pm$	0.06 	&	7.92 	$\pm$	0.12 &$-$	&	11.55$\pm$0.09	&	$-$&SB	\\
CGCG 052-037        	&	8.15 	$\pm$	0.08 	&	8.19 	$\pm$	0.06 	&	8.19 	$\pm$	0.08 &$-$	&	11.38$\pm$0.09	&	$-$&SB	\\
IRAS F16399-093      	&	8.28 	$\pm$	0.06 	&	8.23 	$\pm$	0.06 	&	7.98 	$\pm$	0.13 &$-$	&	11.56$\pm$0.09	&	$-$&SB	\\
NGC 6285             	&$<	7.65 $			&	$<7.49 	$		&	$<7.64 	 $	&$-$&	11.56$\pm$0.09	&	$-$&SB	\\
NGC 6286            	&	8.31 	$\pm$	0.06 	&	8.32 	$\pm$	0.05 	&	$<7.70 	$	&$-$	&	11.32$\pm$0.09	&	$-$&SB	\\
IRAS F17138-1017     	&	7.97 	$\pm$	0.05 	&	8.06 	$\pm$	0.05 	&	7.72 	$\pm$	0.10 &$-$	&	11.42$\pm$0.09	&	$-$&SB	\\
UGC 11041            	&	8.16 	$\pm$	0.07 	&	8.16 	$\pm$	0.06 	&	$<7.75 	$	&$-$	&	11.04 $\pm$0.09	&	$-$&SB	\\
CGCG 141-034         	&	8.15 	$\pm$	0.09 	&	$<7.83 	$		&$<	7.98 	$	&$-$	&	11.13$\pm$0.09&	$-$&SB	\\
IRAS 18090+0130     	&	8.26 	$\pm$	0.06 	&	8.33 	$\pm$	0.06 	&	7.88 	$\pm$	0.12 &$-$	&	11.58$\pm$0.09	&	$-$&SB	\\
NGC 6701            	&	8.25 	$\pm$	0.03 	&	8.16 	$\pm$	0.04 	&	7.87 	$\pm$	0.08 &$-$	&	11.05$\pm$0.09	&$-$&	SB	\\
NGC 6786            	&	8.01 	$\pm$	0.07 	&	8.05 	$\pm$	0.09 	&	7.85 	$\pm$	0.11 &$-$	&	11.43$\pm$0.09	&	$-$&SB/AGN	\\
UGC 11415           	&	7.75 	$\pm$	0.00 	&	8.04 	$\pm$	0.10 	&$<	7.65 	$	&$-$	&	11.43$\pm$0.09	&	$-$&SB	\\
ESO 593-IG 008      	&$<	8.83 	$		&$<	8.92 	$		&$<	8.86 	$	&$-$	&	11.87$\pm$0.09	&	$-$&SB	\\
NGC 6907             	&	7.97 	$\pm$	0.06 	&	7.71 	$\pm$	0.09 	&	7.42 	$\pm$	0.07	&$-$	&	11.03$\pm$0.09	&$-$&	SB	\\
IRAS 21101+5810     	&	8.37 	$\pm$	0.07 	&	8.17 	$\pm$	0.11 	&	7.99 	$\pm$	0.01	&$-$	&	11.75$\pm$0.09	&	$-$&SB	\\
ESO 602-G 025       	&	8.20 	$\pm$	0.11 	&	8.16 	$\pm$	0.07 	&	8.15 	$\pm$	0.12 &$-$	&	11.27$\pm$0.09	&	$-$&SB/AGN	\\
UGC 12150           	&	8.28 	$\pm$	0.05 	&	8.35 	$\pm$	0.04 	&	7.86 	$\pm$	0.12 &$-$	&	11.29$\pm$0.09	&	$-$&SB	\\
IRAS F22491-1808    	&	$<9.09 	$		&	$<9.04 	$		&	$<9.03 	$	&$-$&	12.11$\pm$0.09	&	$-$&SB	\\
CGCG 453-062        	&	8.26 	$\pm$	0.07 	&	7.95 	$\pm$	0.14 	&$<	7.90 	$	&$-$	&	11.31$\pm$0.09	&	$-$&SB	\\
NGC 7591            	&	8.06 	$\pm$	0.06 	&	7.76 	$\pm$	0.08 	&	7.76 	$\pm$	0.13 &$-$	&	11.05$\pm$0.09	&	$-$&SB/AGN	\\
IRAS F23365+3604    	&	8.94 	$\pm$	0.14 	&	9.14 	$\pm$	0.09 	&	$<8.98 	$	&$-$	&	12.13$\pm$0.09	&	$-$&SB/AGN	\\
\hline

\hline	

 		

\end{longtable}
{Notes. $^{\mathrm{a}}$ The measured the IR emission within the beam-size of 28$''$. 
$^{\mathrm{b}}$} The measured the IR emission within the beam-size of 17$''$. 
For the sources in \cite{2015ApJ...814...39P}, the infrared luminosity is from \cite{2003AJ....126.1607S} and the uncertainties of L$_{\rm IR}$ are assumed to be 20\%.

\clearpage
\twocolumn









\clearpage

\appendix

\section{Spectra of dense gas tracers for individual galaxies}


\begin{figure*} 
    \centering
   \includegraphics[height=9in,width=6.4in]{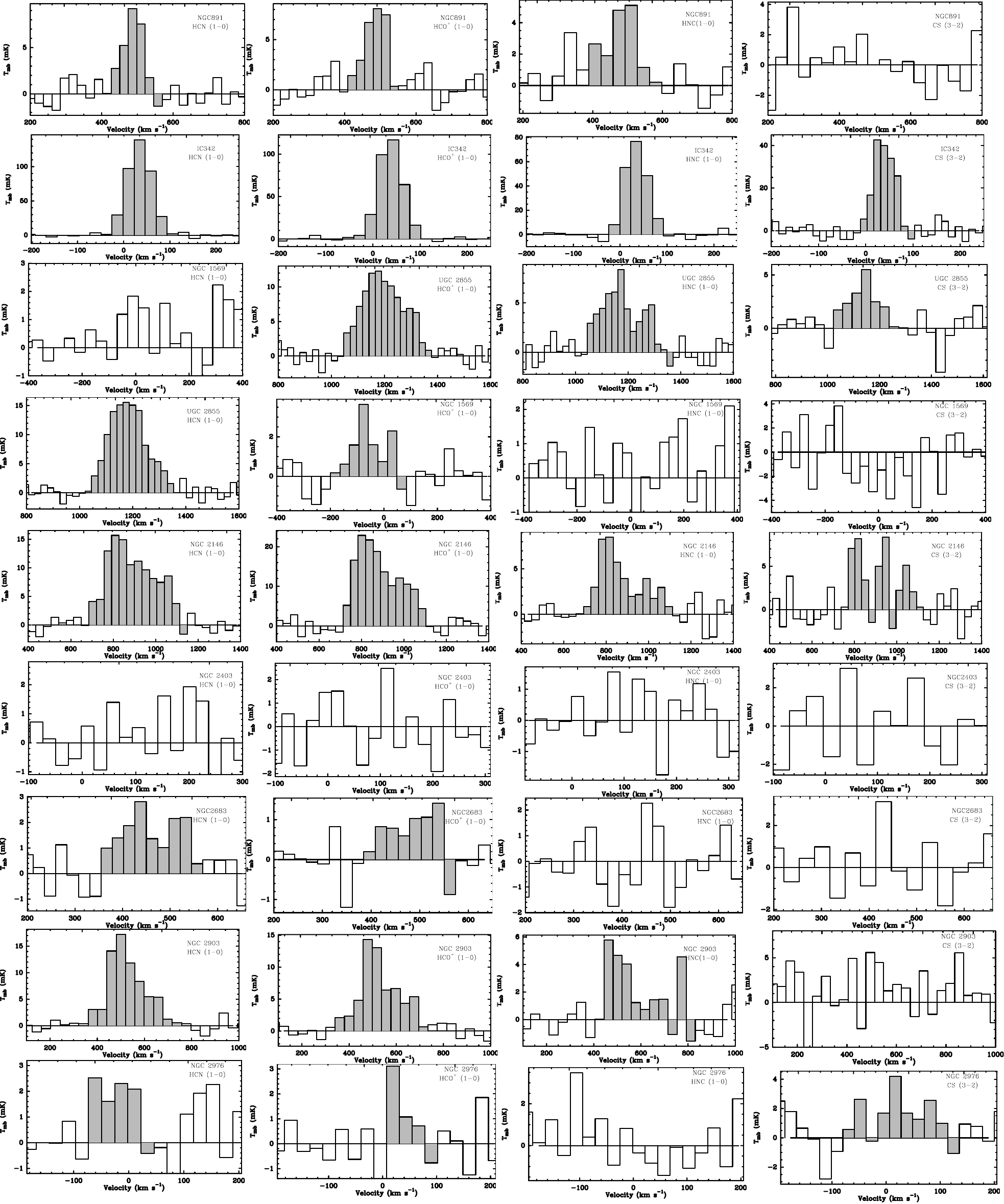}
    \caption{The grey-filled black spectra shown in each subfigure are the detected lines, otherwise, the spectra just shown the marginal detection or the non-detection of those molecular transitions. }
    \label{fig:f1}
\end{figure*}

\begin{figure*} 
    \centering
    \includegraphics[height=9in,width=6.4in]{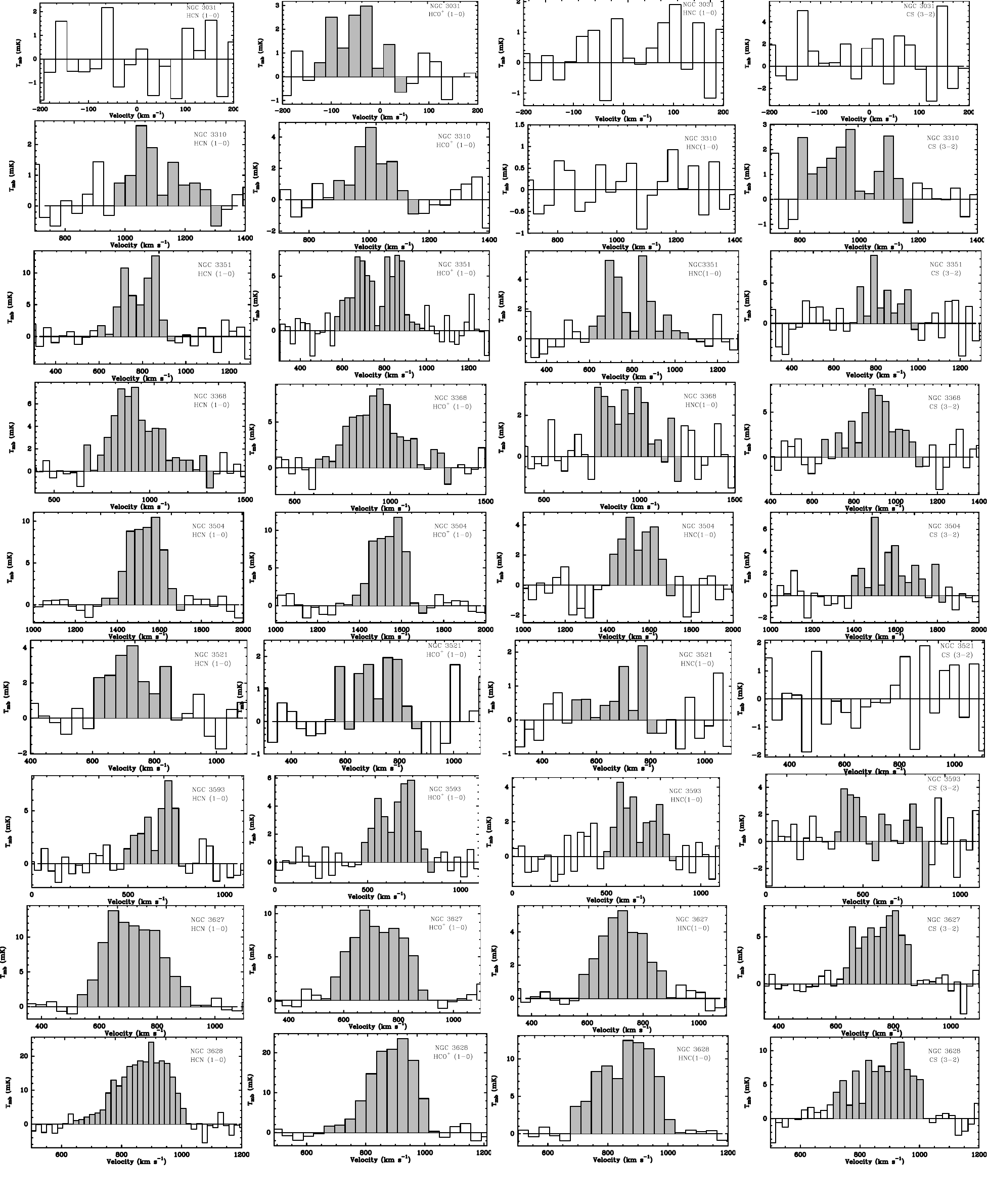}
    \contcaption{}
    \label{fig:f2}
\end{figure*}
\begin{figure*} 
    \centering
    \includegraphics[height=9in,width=6.4in]{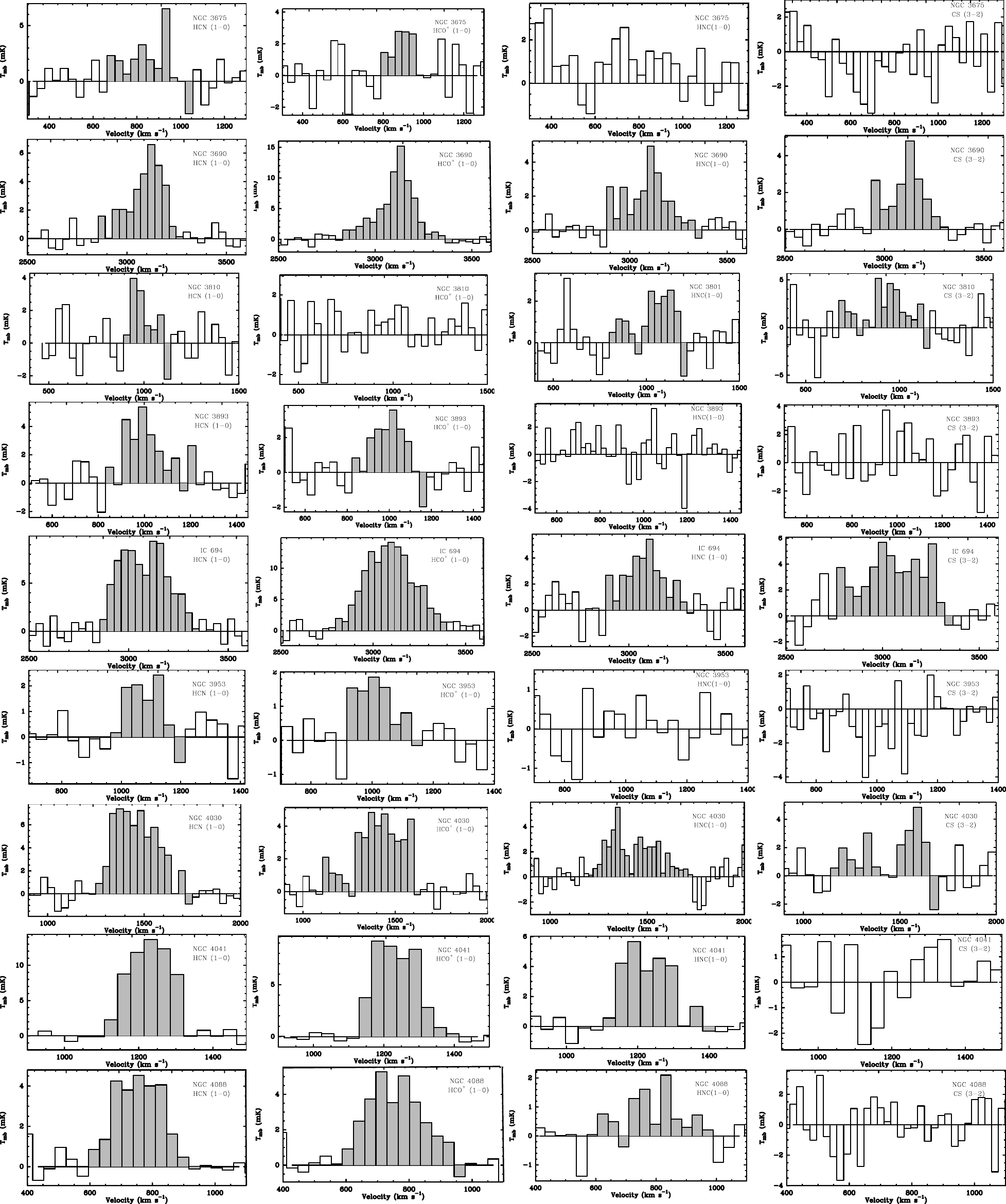}
    \contcaption{}
    \label{fig:f3}
\end{figure*}

\begin{figure*} 
    \centering
    \includegraphics[height=9in,width=6.4in]{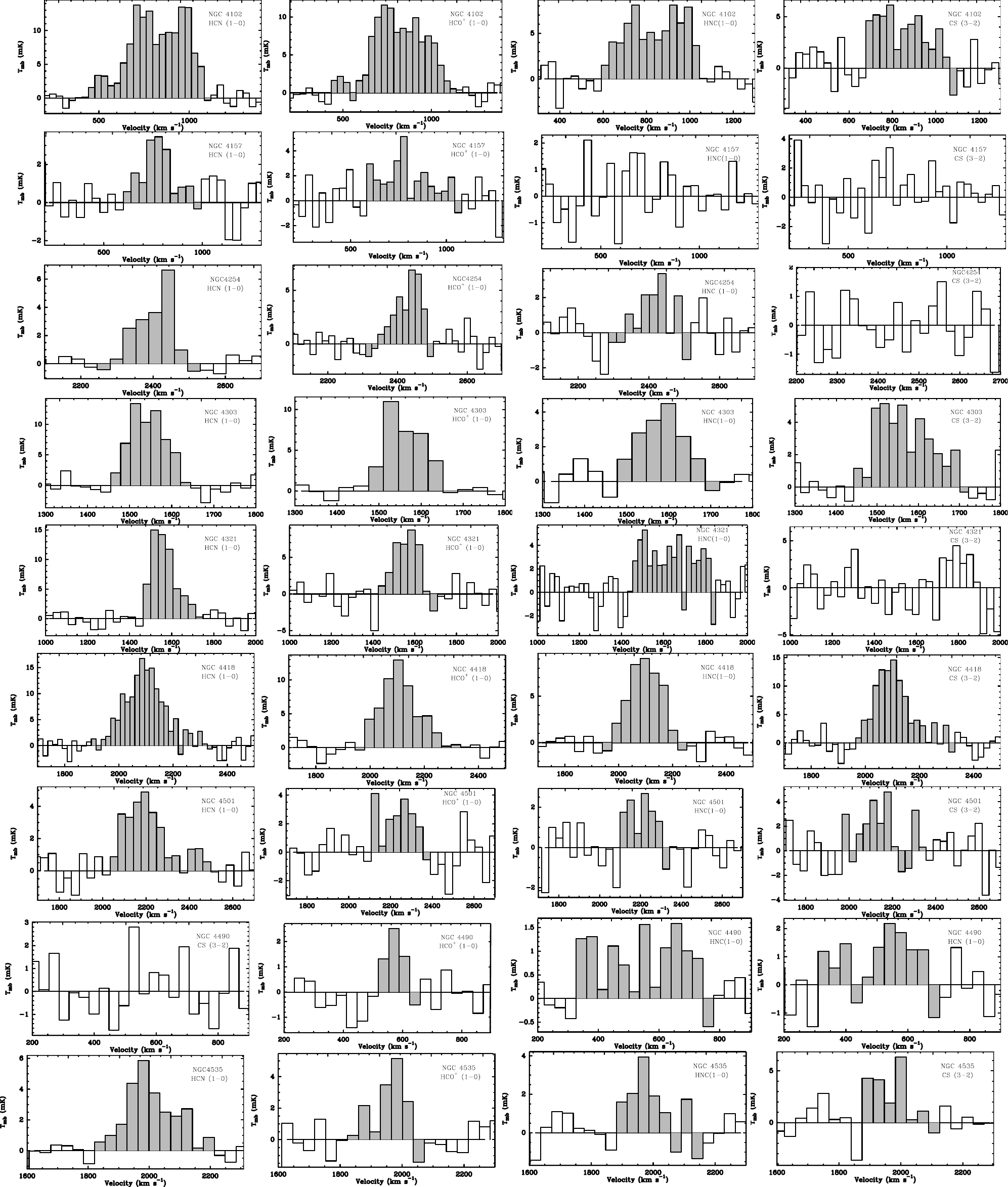}
    \contcaption{}
    \label{fig:f4}
\end{figure*}

\begin{figure*} 
    \centering
    \includegraphics[height=10in,width=6.4in]{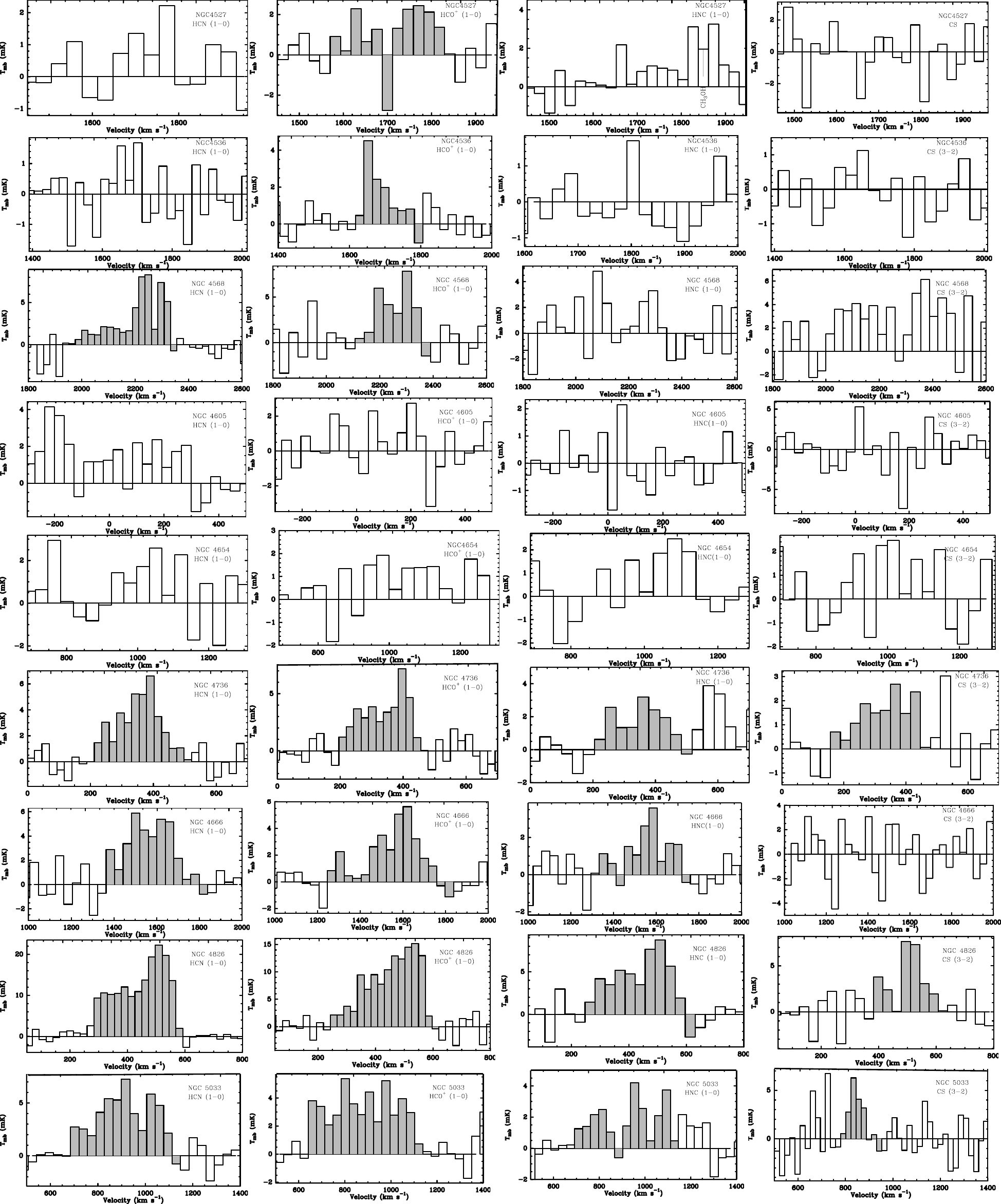}
    \contcaption{}
    \label{fig:f5}
\end{figure*}

\begin{figure*} 
    \centering
    \includegraphics[height=9in,width=6.4in]{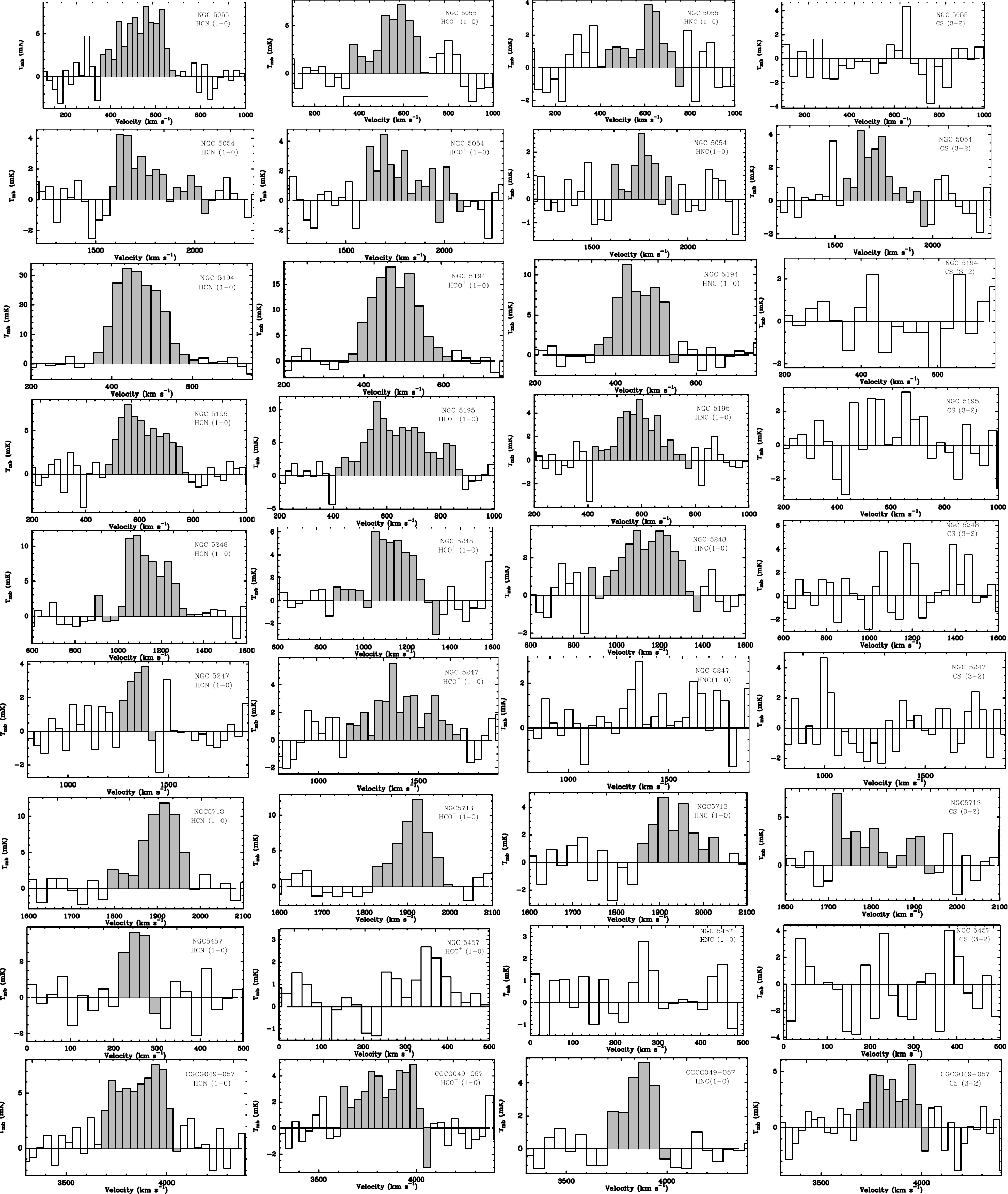}
    \contcaption{}
    \label{fig:f6}
\end{figure*}

\begin{figure*} 
    \centering
    \includegraphics[height=4in,width=6.4in]{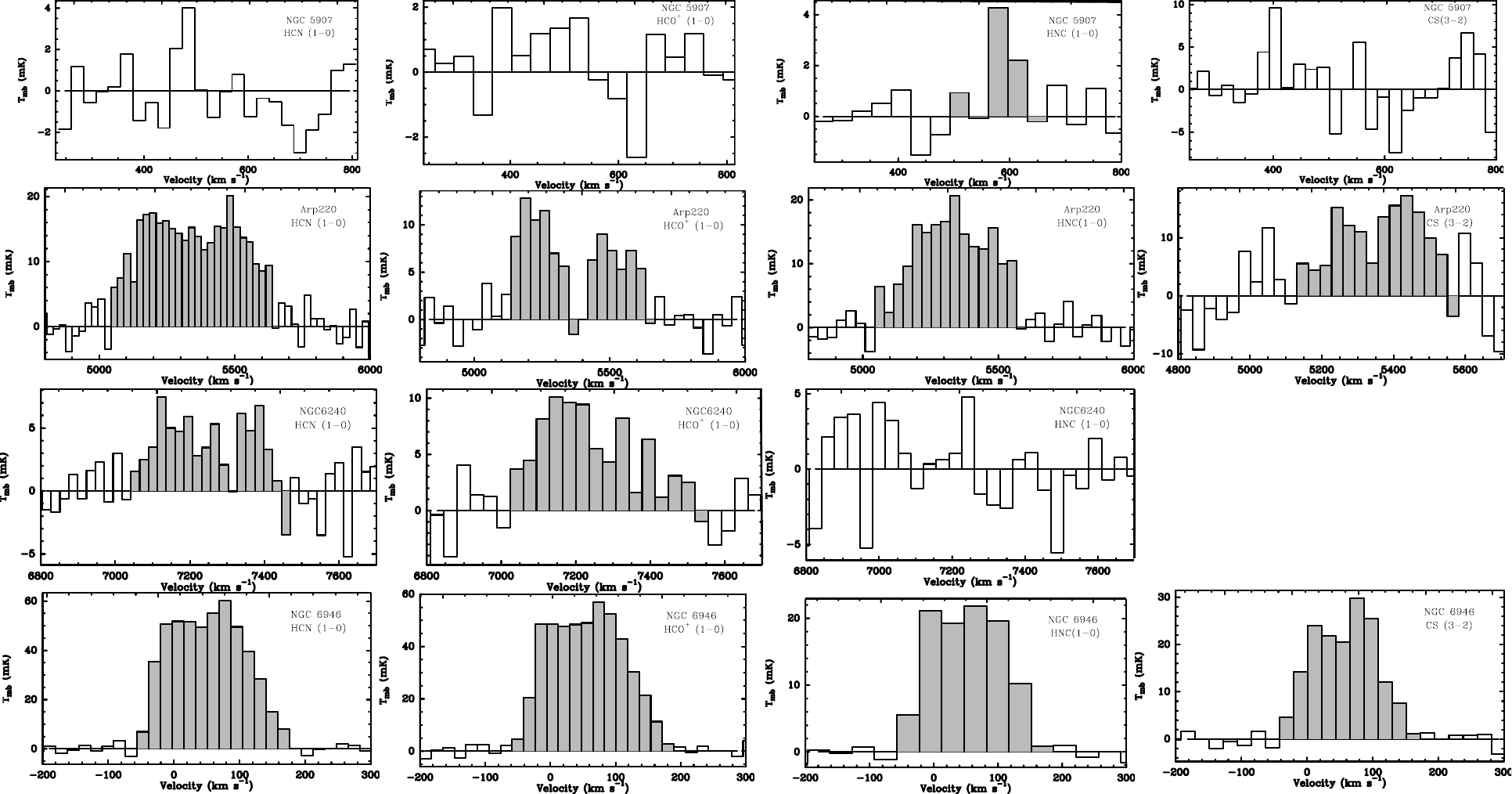}
    \contcaption{}
    \label{fig:f7}
\end{figure*}


\bsp	

\label{lastpage}

\end{document}